\documentclass[11pt,a4paper]{article}
\usepackage[titletoc]{appendix}
\usepackage{bbm}
\usepackage{natbib}
\usepackage{charter}
\usepackage{pgf}
\usepackage{tikz}
\usetikzlibrary{arrows,automata}

\usepackage{verbatim}
\usepackage{subcaption}
\usepackage[left=2cm,top=3.5cm,bottom=3.5cm,right=2cm]{geometry}
\usepackage[utf8]{inputenc}
\usepackage[T1]{fontenc}
\usetikzlibrary{math}

\usepackage{amssymb,amsthm,amsmath,xcolor}

\newtheorem{theorem}{Theorem}
\newtheorem{proposition}{Proposition}
\newtheorem{lemma}{Lemma}[section]
\newtheorem{definition}{Definition}

\newtheorem{corollary}{Corollary}
\newtheorem{remark}{Remark}

\newtheorem{example}{Example}

\title{Revealed Social Networks\thanks{We are thankful to the editor, Paola Manzini, and two anonymous referees for comments that helped improve the paper. We are also thankful to Matteo Bizzarri, Andreas Bjerre-Nielsen, Luca Braghieri, Tugce Cuhadaroglu, and Fernando Vega-Redondo as well as seminar participants at the Italian Junior Workshop on Economic Theory and NSEC 2025 for their helpful comments and discussions.}}

\author{Christopher P. Chambers\thanks{Department of Economics, Georgetown University, ICC 580  37th and O Streets NW, Washington DC 20057. E-mail: \texttt{Christopher.Chambers (at) georgetown.edu}.} \and Yusufcan Masatlioglu\thanks{Department of Economics, University of Maryland, 3147E Tydings Hall, 7343 Preinkert Dr.,  College Park, MD 20742. E-mail: \texttt{yusufcan (at) umd.edu.}} \and Christopher Turansick\thanks{Department of Decision Sciences, Bocconi University and IGIER, Via Roentgen, 1, Milan, Italy 20136.  E-mail:  \texttt{christopher.turansick (at) unibocconi.it} }}

\date{May 15, 2026}

\begin{document}

\maketitle

\begin{abstract}
The linear-in-means model is the standard empirical model of peer effects and asks that an agent's choice or outcome is a combination of their ideal point and the mean outcome of their group. Using choice data and exogenous group variation, we develop a revealed preference style test for the linear-in-means model. This test is formulated as a linear program and can be interpreted as a condition about differentiating the behavior of each agent in a consistent manner. We then study the identification properties of the linear-in-means model. A key takeaway from our analysis is the close relationship between the dimension of the outcome variable and identification. When the outcome variable is one-dimensional, failures of identification are generic. When the outcome variable is multi-dimensional, we provide natural conditions under which identification is generic.

\end{abstract}
\section{Introduction}
People rarely make decisions in isolation and are often influenced by their neighbors and peers.\footnote{Social influence has been studied in an array of contexts including test scores \citep{sacerdote2001peer}, worker productivity \citep{mas2009peers}, alcohol use \citep{kremer2008peer}, risky behavior \citep{card2013peer}, and tax compliance \citep{fortin2007tax}.} This influence can take the form of a social norm, a common convention, or conformism. The linear-in-means model serves as a foundational framework for estimating peer effects in economics and social science research. The model supposes that an agent's choice is a convex combination of their own ideal point and the weighted average of each other agent's choice. The objective of this paper is to investigate the behavioral implications of the linear-in-means model leveraging revealed preference techniques \`a la Afriat. Moreover, we establish sufficient conditions to jointly recover each agent’s preferences and the underlying social network, including the strength of every connection.

As \citet{manski1993identification} emphatically observed, the identification of social influence parameters has long been recognized as a challenging task. The crux of the reflection problem lies in disentangling and separately identifying exogenous group effects from endogenous peer effects. In this paper, we tackle the identification of network effects through the lens of stochastic discrete choice theory—an approach that yields several novel, intuitive, and surprisingly powerful insights.

We consider a choice procedure based on the model of \citet{manski1993identification} wherein each agent chooses an action which is a convex combination of their ideal point and a weighted average of the other agents' actions.\footnote{Our base specification restricts to the case of convex weights, but we extend to more general linear combinations in Section \ref{sec:negweights}.} Formally,  
\begin{equation*}
    p_i^N=\pi_i^N(i) v_i + \sum_{j \in N \setminus i}\pi_{i}^N(j) p_j^N
\end{equation*} where the vectors of $v_i$ and $p_i^N$ correspond the agent $i$'s ideal point and the action taken by agent $i$ in group $N$, respectively.  Further, $\pi_i^N(j)\geq 0$  denotes the impact of agent $j$'s action on agent $i$ in the context of group $N$. While $p_i^N$s are observable, $v_i$ and $\pi_i^N$s are not observable and need to be identified from observed choices.

Examples of this model are in abundance. Schools collect data on how often each student participates in different after-school activities such as sports, music, and study clubs. Each student spreads their time across these categories (e.g., some weeks they play soccer more, other weeks they practice violin). The same students might appear in different groups—sometimes two friends share a class, sometimes they do not. Although their observed activity frequencies change depending on the group they are in, the pattern of who influences whom stays the same. In a corporate setting, employees log the fraction of their weekly time spent on meetings, project work, and administrative tasks. Employees often work in rotating teams. Their observed time allocations shift depending on which coworkers are present in a given week. However, their underlying influence network—for example, whose work habits they emulate—remains stable across all task categories. Households vary their shopping baskets over time, allocating spending across categories like fresh produce, snacks, and beverages. The recorded grocery data reflects stochastic distributions over categories. Families also differ in which neighbors they regularly interact with—for example, those who exchange recipes or dietary advice. These interpersonal influences persist across all food categories: the neighbors someone pays attention to when choosing produce are the same neighbors influencing their beverage choices.

We now illustrate how to do identification in a simple example. Consider three college friends—Ann, Ben, and Can—and their sports choices. While Can was studying abroad, Ann chose to play tennis 50\% of time, volleyball 10\% of the time, and walk 40\% of the time, while Ben's frequencies were 70\%, 10\%, and 20\%, respectively. This data alone does not solve the endogeneity problem. An observer cannot discern whether behavioral similarities arise from peer influence or similar preferences/backgrounds that would produce identical choices even without interaction. These explanations remain observationally equivalent without further data. Our approach resolves this by leveraging natural group variation. For instance, suppose that Can returns but Ben leaves to study abroad the following semester. Now, we observe that Ann's choices are tennis (10\%), volleyball (50\%), and walking (40\%), while Can's frequencies were 10\%, 70\%, and 20\%, respectively. By comparing choice behavior across the two periods (the first semester choices of Ann and Ben without Can and the second semester choices of Ann and Can without Ben) we can uniquely isolate peer effects (point identification), provided the observed choices satisfy our model's assumptions (formalized later).\footnote{In this example, our unique identification reveals that Ann's favorite exercise is walking, even though it is not the most frequently chosen by Ann. Indeed, her ideal point is $v_{Ann}=(0.1,0.1,0.8)$. On the other hand, tennis and volleyball are Ben's and Can's favorite sports, respectively. The data further show that Ann weighs her friends' choices twice as heavily as her own ideal point (a 2:1 ratio).}

Our identification strategy relies on (1) group variation $\{$Ann, Ben$\}$ vs $\{$Ann, Can$\}$ and (2) multi-dimensional choice objects (tennis, volleyball, walking).\footnote{In this paper, our main form of variation is group/participation variation. However, our main results can be applied in a setting where there is no group variation but there is variation of some observable variable that is a perfect instrument for variation in the underlying network structure of a fixed group.}\footnote{We note that group variation shows up in various empirical and quasi-experimental settings. Variation in an agent's group can arise via intentional randomization in experiments \citep{sacerdote2001peer,falk2006clean}, varying classroom composition \citep{duflo2011peer}, or varying team or shift composition at work \citep{mas2009peers}.} A key insight of our analysis is that a necessary condition for recovery of the underlying network structure is that the number of agents in each of our groups must be no more than the dimension of our outcome variable. We emphasize that our identification approach does not depend on assumptions about (i) how social networks differ across groups, (ii) what observable features characterize the activities, or (iii) what preferences individuals hold. This distinguishes our contribution and complements existing work that requires these assumptions.\footnote{Much of the work on identifying parameters in the linear-in-means model assumes (partial) observability of the underlying network structure \citep{bramoulle2009identification,blume2015linear}. Recently,  \citet{lewbel2023social} and \citet{de2024identifying} provide sufficient conditions under which the underlying network structure can be recovered without data on the network itself. Unlike ours, these papers require enough variations on observable characteristics. Since our approach is complementary to these papers, one might allow a more general model where point identification is possible under weaker conditions.}

Given that our identification results rely on the particular choice procedure we have adopted, it is natural to consider the falsifiability of this model. We show that our model is fully characterized by a linear program condition in the spirit of \citet{afriat1967construction}. This result renders our model behaviorally testable without any restrictions across groups or about the weights assigned to each agent. Our test characterizes datasets of the previously described form which are consistent with the linear-in-means model via an easily solvable linear program. We interpret this linear program as asking that, whenever an agent's behavior differentiates the agent from their group, the direction it differentiates the agent must be consistent across each of the agent's groups.\footnote{Testing models of peer effects is difficult as many people choose peers who are similar to them in observable characteristics. As a result of this difficulty, many studies that aim to test models of peer effects, including the linear-in-means model, do so through natural/quasi/pure experimental methods \citep{sacerdote2014experimental,basse2024randomization}. In addition, experimental methods are also often used when quantifying the impact of peer effects. See \citet{agranov2021importance} as an example.}

Thus far, we have made no assumptions about the underlying social network structure across groups. While this is a strong point of our testing and identification results, this poses a problem for the predictive power of our model. To solve this problem, we consider a refinement of the linear-in-means model, which assumes a common social network structure across groups. The strength of network connections from one person to another only varies across groups due to renormalization. Under this assumption, we provide sufficient conditions that allow us to predict choices in any possible group. To test the validity of this assumption, we develop an extension of our testing procedure for the general case. This version of the linear-in-means model is characterized by a strengthening of our previous linear program. The updated version asks for consistency in the way that an agent's behavior differentiates itself from the across-group average behavior of any other agent.

Finally, we consider a version of the linear-in-means model more in line with the original reflection problem posed by \citet{manski1993identification}. In this version of the model, a person is influenced by each other person uniformly. This corresponds to the unweighted average choice of every other person in the group being the social norm faced by an agent. Alternatively, this setup corresponds to when agents face (linear) peer effects but do not face more complicated forms of network effects. In this case, we can once again strengthen our original linear program in order to test for the absence of network effects in the context of the linear-in-means model. This linear program asks for consistency across groups in the way that an agent's behavior differentiates itself from the within-group average behavior of each other agent.

The rest of this paper is organized as follows. In Section \ref{Sec:Model}, we formally introduce the linear-in-means model and our notation. In Section \ref{Sec:Results}, we introduce and discuss our testing and identification results for the three different specifications of the linear-in-means model.  In Section \ref{Sec:Extensions}, we discuss how our results extend to alternative types of data, under weaker assumptions on how peers influence each other, and to the case when there are more agents than outcome dimensions. Finally, we conclude with a discussion of the related literature in Section \ref{Sec:Discussion}.

\section{Model and Preliminaries}\label{Sec:Model}
Our interest is in studying the linear-in-means model of social interaction. We build on the base model in two meaningful ways. First, instead of restricting an agent's choice to be from an interval, we allow agents to choose a distribution over a finite set of goods or outcomes. Second, we consider stochastic choice data that arises when the set of agents present varies.

\subsection{Preliminaries and Notation}
Denote by $\mathcal{A}$ the grand set of agents. Assume that $|\mathcal{A}| \geq 2$. A typical group of agents will be denoted $N$, where $\varnothing\neq N \subseteq \mathcal{A}$.  We let $\mathcal{N}\subseteq 2^{\mathcal{A}}$ be any set of \emph{groups}. For any agent $i\in \mathcal{A}$, let $\mathcal{N}_i=\{N\in\mathcal{N}:i\in N\}$ denote the set of groups to which $i$ belongs. We sometimes abuse notation and use $N \setminus i$ to denote the group of agents formed by removing agent $i$ from group $N$. Let $X$ be some finite set of alternatives or outcomes. Enumerate these alternatives from $1$ to $|X|$.  We use $(0,\dots,0,1,0,\dots,0)$, where $1$ is in the $n$th dimension, to denote the $n$th alternative and use $x,y \in X$ to denote arbitrary alternatives. The data in our model consists of, for each $N\in\mathcal{N}$ and each $i\in N$, a probability distribution over $X$. Formally, for $i\in N\in\mathcal{N}$, this is denoted $p_i^N\in\Delta(X)$. We use $p_i^N(x)$ to denote the choice probability of good $x$ by agent $i$ in group $N$ and $p^N$ to denote the matrix where each row corresponds to $p_i^N$ for a different $i \in N$. We will sometimes use $p_{-i}^N$ to denote the matrix which is formed by taking $p^N$ and removing the row corresponding to agent $i$.

Before moving forward, we note that, while our focus is on probabilities, all of our results in Section \ref{Sec:Results} can be suitably adapted to working with data that comes from any compact and polyhedral subset of Euclidean space. As an example, we may consider as data people's SAT scores broken down into their scores on the reading section, the writing section, and the math section. We discuss the extension of our main results to this more general Euclidean setup in Section \ref{Sec:Extensions}.

\subsection{The Model}

In the linear-in-means model, each agent's action is a mixture of their own ideal point and the actions of others. We use $v_i \in \Delta(X)$ to denote agent $i$'s ideal point. This corresponds to the action agent $i$ would take in isolation. We introduced $p_i^N$ as our data, but it also corresponds to the action taken by agent $i$ in group $N$. The amount that agent $j$'s action impacts agent $i$'s action may differ from the amount that agent $k$'s action impacts agent $i$'s action. In fact, the impact of agent $j$'s action on agent $i$ may depend on the context or the group of agents currently present. We use $\pi_i^N(j)\geq 0$ to denote the impact of agent $j$'s action on agent $i$ in the context of group $N$. Similarly, we use $\pi_i^N(i)>0$ to denote the impact of agent $i$'s ideal point on agent $i$ in group $N$. We assume that $\sum_{j \in N}\pi_{i}^N(j)=1$. These impact or influence weights enter into agent $i$'s action in a linear manner.

\begin{equation}\label{EQ:linInMeans}
    p_i^N=\pi_i^N(i) v_i + \sum_{j \in N \setminus i}\pi_{i}^N(j) p_j^N
\end{equation}

\noindent Equation \ref{EQ:linInMeans} tells us that agent $i$'s action in group $N$ is given as a convex combination of their ideal point and a weighted average of the other agents' actions. We now discuss three key modeling assumptions which differ from the most general version of the linear-in-means model.

Our first assumption is that an agent's outcome is restricted to the simplex. This means that the total value of agent $i$'s outcome and the total value of agent $j$'s outcome are restricted to be the same. This rules out situations where the total value of each agent's outcome can vary, either across groups or across agents.\footnote{One such example of a situation is the case of exam scores. Suppose each dimension corresponds to an agent's score on an exam in a specific subject. If we want to allow for the sum of scores to vary across agents or groups, this simplex assumption is a meaningful restriction.} In Section \ref{Sec:Extensions}, we discuss how our main results on testing and identification extend to the case when outcomes are allowed to be anywhere in a convex subset of Euclidean space.

Our second assumption is that $\pi_i^N \geq 0$ and $\sum_{j \in N} \pi_i^N(j) =1$. In other words, we assume that the total influence an agent faces, including their impact on themselves, is non-negative and constant across groups.\footnote{This assumption on the linear-in-means model is partially due to our data assumptions. The fact that each $p_i^N$ lies in the simplex necessitates $\sum_{j \in N} \pi_i^N(j)=1$. If this were not the case, our data would lie outside the simplex.} Under this assumption, the linear-in-means model is best interpreted as a model of conformism. Our agent is influenced by the weighted average of the other agents' choices and has some incentive to match this average. In Section \ref{Sec:Extensions}, we discuss how we can extend our main results on both testing and identification to the case when weights are allowed to be negative and not sum to one, so long as $\pi_i^N(i)\in (0,1]$.

Our final assumption and our key assumption concerns how $\pi_i$ behaves across the dimensions of the choice problem. We impose that the influence parameters $\pi_i^N(j)$ are \emph{identical in every dimension}. In our running example, Ann responds to Ben and Can's choices in the same way regardless of whether the choice concerns tennis, volleyball, or walking. This restriction is innocuous in the one-dimensional case, but is meaningful in the multidimensional case. One could imagine the cases where the importance of agent $j$ to agent $i$ varies by dimension. For instance, the influence of Ben and Can can vary in choices regarding travel, food, or sports. Our assumption can be interpreted as focusing on a subset of dimensions for which influence is relatively constant. This restriction is essential for our main results and cannot be relaxed.

While this assumption is not applicable to every environment (see \citet{belhaj2014competing} and \citet{zenou2024games}), we argue that this assumption is reasonable in certain cases. As a first example, we suggest there may be certain sets of choices that share the same network. Consider the choices of professors in a department. When considering how to split their time between research, teaching, and departmental service, it is reasonable to expect that one professor's influence on another is constant across these dimensions. As a second example, we suggest that this assumption may be more reasonable for agents in certain age ranges (or more generally, specific groups). Our prime example is the choice of children or teenagers who have yet to build distinct social networks (i.e. work colleagues, research networks, in-laws, etc.) and are mostly impacted by their school friends.

A key part of our analysis is observing data across various groups. As there are no restrictions on who is present in each of these groups, the underlying network structure can vary from group to group. This group and network variation can be modeled through different assumptions on $\pi_i^N$. Intuitively, $\pi^N$ is a matrix that captures weighted directed influence. By assuming that $\sum_{j \in N} \pi_i^N(j)=1$, we are assuming that $\pi^N$ is a stochastic matrix. When we observe group variation, each group $N$ is subject to its own $\pi^N$. To demonstrate the richness of our framework and motivate the analysis to follow, we discuss seven examples of linear-in-means models. 

\begin{enumerate}

 \item \textbf{The Uniform Model (ULM):} This is the simplest linear-in-means model. An agent cares about each agent, other than themself, in group $N$ equally. That is, for each agent $i$ and each group $N \in \mathcal{N}_i$, there is an $\alpha^i_N \in (0,1]$, such that $\pi_i^N(i)=\alpha_i^N$ and, for each $j\in N$, $$\pi_i^N(j)=(1-\alpha_i^N)\frac{1}{|N|-1}$$ (\emph{i.e.} an unweighted average). This corresponds to the case where there are no network effects and agents influence each other purely through their impact on the group's average outcome. This specification of the linear-in-means model corresponds to the model considered in \citet{manski1993identification} without attributes.
 
 \item \textbf{The Luce Model (LLM):} This model captures a non-symmetric complete network model. Each agent $i$ can be thought of as having, for each agent $j$, an invariant importance weight $w_i(j)$. The relative importance of agent $j$ to agent $i$ in population $N$ is simply the renormalization of each of these importance weights so that they sum to one. Formally, for each $i,j\in \mathcal{A}$, there is $w_i(j)$ such that for all $i\in\mathcal{A}$, $w_i(i)>0$, and $\sum_{j \in \mathcal{A}} w_i(j)=1$ such that $$\pi_i^N(j)=\frac{w_i(j)}{\sum_{k \in N} w_i(k)}$$

More simply, LLM can be thought of as modeling the existence of a single underlying network that is group invariant. Choice in the context of group $N$ corresponds to renormalizing the fixed network weights.

  \item \textbf{The Club Model:} This model aims to capture the differences in peer effects for different groups. We have in mind studying, for example, the peer effects of belonging to different clubs/cohorts/dorms. Let $P$ be a partition of $\mathcal{A}$ where $P=\{P_1,\dots,P_n\}$. $P$ represents a collection of disjoint cohorts/clubs, where $P_k$ represents all the people in club $k$. Each agent is influenced by each club differently. Let $\alpha_{ik}$ be agent $i$'s weight for club $P_k$, where $\alpha_{ik}$ is a parameter capturing the popularity of club $k$ for club member $i$. Within the club, the influence is symmetric, as in ULM.\footnote{One could generalize this example by assuming within-club influence as modeled in LLM.} Formally, for each $i,j\in N$ such that $j \in P_k$, $$\pi_i^N(j)=\frac{\alpha_{ik}}{\sum\limits_{P_l\cap N \neq \emptyset}\alpha_{il}}\frac{1}{|P_k \cap N|}$$

         \item \textbf{The Friendship Model:} This model allows us to capture peer effects based on a ranking of friends. The weight of each friend depends on the relative ordering of the friend in the group. For example, an agent's best friend may be assigned the highest influence factor.  Let $R_i$ (complete and transitive) be a ranking of the agents other than $i$, representing friendship. Agent $i$ assigns weight according to the ranking of the other agents. An agent puts a weight of $\alpha_i \in (0,1)$ on her ideal point, and then cares about the other agents according to their ranking. Agent $i$ assigns weight $w(j,r_i(j,N))$ to agent $j$ where $r_i(j,N)$ is the ranking of $j$ in $N$ in according to $R_i$.  For each $i,j\in N$, $$\pi_i^N(j)=(1-\alpha_i) \frac{w(j,r_i(j,N))}{\sum_{k \in N} w(k,r_i(k,N))}$$

\item \textbf{The Random Participation Model:} This model aims to capture the classical disconnect between analyst and agents. The analyst observes group data, such as a class roster, but is unable to observe the actual participation in that group, such as the attendance record for each agent. Each agent's attendance is random, with the probability of agent $i$'s attending the course given by $\gamma(i) \in (0,1)$. The agents face a fixed underlying network, which is renormalized based on the realization of attendance. The weights are proportional to the amount of time spent by that agent in the same classroom.\footnote{One could think of a version of the random participation model where participation is correlated across agents and is group dependent. In this case, $\phi(M,N) \in [0,1]$, with $\sum_{\emptyset \subsetneq M \subseteq N}\phi(M,N)=1$, represents the probability of $M$ being the realized group when $N$ is the observed group. This leads to the following model.$\pi_i^N(j)=\sum\limits_{i \in M \subseteq N} \phi(M,N) \frac{w_i(j)}{\sum\limits_{m \in M}w_i(m)}
$.}

\begin{equation*}
    \pi_i^N(j)=\sum\limits_{i \in M \subseteq N} \prod_{k \in M \setminus i} \gamma(k) \prod_{l \in N \setminus M} (1- \gamma(l)) \frac{w_i(j)}{\sum\limits_{m \in M}w_i(m)}
\end{equation*}
\noindent

% \item \textbf{The Random Arrival Model:} Think of a group of friends who arrive at the cafeteria at different times. This creates a different seating arrangement every time. Over the course of repeated interactions, agents face a distribution over these arrangements. The impact of one agent on another corresponds to a mixture of their impact across various fixed networks/arrangements. Let $q(A)$ be the probability of arrangement $A$.

% \begin{equation*}
%     \pi_i^N(j)=\sum\limits_{A} q(A) \frac{w_A(j)}{\sum\limits_{k \in N}w_A(k)}
% \end{equation*}

  \item \textbf{The General Model (GLM):} $\pi_i^N(j)$ is not assumed to satisfy any hypotheses across $N$ except $\pi_i^N(i)>0$.\footnote{In Appendix \ref{app:GLMExtension} we consider the case when $\pi_i^N(i)$ is only assumed to satisfy non-negativity and extend all of our results to this case.} The relative importance of each agent $j$ to agent $i$ is allowed to be fully context/group dependent. The interpretation here is that an agent $j$ could act as a complement or substitute for agent $k$ in the context of a group.\footnote{Consider a setting where agent $j$ has some noisy information about an underlying state. Agent $i$ finds it important to match their action with agent $j$ in order to coordinate with the underlying state. Now suppose that agent $k$ is added to the group. Agent $k$ has perfect knowledge of the underlying state. In this case, the relative importance of agent $j$ to agent $i$ would go to zero when agent $k$ is introduced as the action of agent $k$ is a better signal for the underlying state.}

\end{enumerate}

Each model in the list above is interesting in its own right. However, due to space constraints, a comprehensive study of all of them is infeasible. Instead, we focus on the extreme cases-- namely, the ULM, LLM, and GLM. Our analysis of the most general form of the GLM will provide insights into identification across these models. That said, each model benefits from additional identification power due to its specific structural assumptions. We therefore encourage future research to explore these models in greater depth. 

Our goal in this paper is to study data that arises from the ULM, LLM, and GLM versions of the linear-in-means model. Our focus is on testing these models and identifying the unobserved parameters of the model ($v_i$ and $\pi_N$). With this in mind, we introduce our definition of consistency.

\begin{definition}
    We say that a dataset $\{p^N\}_{N \in \mathcal{N}}$ is \textbf{consistent} with GLM/LLM/ULM if there exists $v_i \in \Delta (X)$ for each $i \in \mathcal{A}$ and $\pi^N$ satisfying the conditions of GLM/LLM/ULM such that Equation \ref{EQ:linInMeans} holds for each $N \in \mathcal{N}$ and $i \in N$.
\end{definition}
\noindent Note that these three models are nested: ULM $\subset$ LLM $\subset$ GLM. 

\subsection{Data and Model Foundations}\label{sec:Foundations}

In this section, we discuss  foundations and interpretation for the linear-in-means model as well as potential data-generating procedures under these foundations. In the standard setup of the linear-in-means model, when $|X|=2$, which we call the one-dimensional case as one dimension is a sufficient statistic for the other, $p_i^N$ is often interpreted as an effort level or a test score. Notably, our focus will be on situations where our data can be multidimensional. Our model allows us to capture higher granularity in choice. As we will see in Section \ref{Sec:Results}, the multi-dimensional aspect of our model allows us to recover generic joint identification of preference and network parameters which generically fails in the one dimension case.

Our first interpretation of the linear-in-means model is as arising as the equilibrium best response of a complete information game. Consider two agents who have preferences over how they split their time between labor, leisure, and volunteering. In addition, these two agents have a desire to spend time with each other. Upon making their preferences known, the agents commit to how they spend their time. In the context of this story, an analyst has access to data on how these two agents have used their time across these three dimensions. Formally, each agent has preferences over the elements of the simplex but faces perturbations of their preferences due to the choices of their peers.\footnote{In this case, an agent can be thought of as having a preference over their choice frequencies rather than repeatedly maximizing (potentially different) static utility functions. This corresponds to an agent who is deliberately stochastic \citep{cerreia2019deliberately} or faces a perturbed utility function \citep{fudenberg2015stochastic} subject to social influence.} Specifically, the choices described in Equation \ref{EQ:linInMeans} arise from the group $N$ playing a complete information game where each agent's utility is given by the following.\footnote{In Equation \ref{EQ:baseutility}, note that if we choose to optimize in each dimension, ignoring across dimension constraints, we are left with $p_i^N(x)$ satisfying Equation \ref{EQ:linInMeans}. Since each dimension is subject to the same $\pi_i^N$, the proposed $p_i^N(x)$ satisfy non-negativity and $\sum_{x \in X} p_i^N(x) =1$. Notably, if we relax the assumption of $\pi_i^N(x)=\pi_i^N(y)$, allowing for non-trivial multiplexing across dimensions, our non-negativity constraints may potentially bind and our formulation would not result from optimization of a quadratic loss utility function. See \citet{belhaj2014competing} for a discussion of this in the two dimension case.}

\begin{equation}\label{EQ:baseutility}
    u(p_i^N,p_{-i}^N)=-\pi_i^N(i) \sum_{x \in X}(p_i^N(x)-v_i(x))^2 - \sum_{j \in N \setminus i} \pi_i^N(j) \sum_{x \in X} (p_i^N(x)-p_j^N(x))^2
\end{equation}

\noindent Under this interpretation, each agent's connection weights $\pi_i^N(j)$ depend on the relative costliness of deviating from agent $j$'s choice or their own bliss point. This is in line with the observation made in \citet{blume2015linear}, \citet{boucher2016some}, \citet{kline2020econometric}, and \citet{ushchev2020social} that the one-dimensional linear-in-means model arises from agents maximizing quadratic loss functions. Under this interpretation of the model, we can think of our data as arising from each agent's choice of time usage. We also note that \citet{golub2020expectations} offers an interpretation of the linear-in-means model as arising from an incomplete information game where, in Equation \ref{EQ:linInMeans}, $v_i$ corresponds to agent $i$'s expectation of some underlying random variable and $p_i^N(j)$ corresponds to agent $i$'s expectation of agent $j$'s action.

Our second interpretation of the linear-in-means model is as the time average of a discrete choice procedure. Consider two agents who repeatedly go out on dinner dates with each other. Each agent has a base random utility function over the food on the menu, but this utility function is perturbed by a preference to conform with their date's expected choice. In this story, an analyst has access to the repeated choices or the long run average of these repeated choices of each agent. Formally, in each period, each agent chooses the alternative which maximizes
\begin{equation}\label{EQ:RUMLinMeans}
    u(x)=\ln\left(\pi_i^N(i) v_i + \sum_{j \in N \setminus i}\pi_{i}^N(j) \mathbb{E}\left[\mathbf{1}\{c_j=x\}\right]\right) + \epsilon_x
\end{equation}
where $c_j$ corresponds to the choice of agent $j$ and $\mathbb{E}$ corresponds to the expected value. We make two further assumptions. First, $\epsilon_x$ is distributed according to the Gumbel distribution, leading to logit choice frequencies \citep{mcfadden1974conditional}.\footnote{Note that the assumption of Gumbel errors can be relaxed given that we jointly change the specification of the utility function.} Second, each agent's beliefs are stationary and correct. This means that $\mathbb{E}\left[\mathbf{1}\{c_j=x\}\right]$ is equal to $p_j^N(x)$.\footnote{While we make no claim about convergence of such a process, this correct expectations assumption can be thought of as arising from an invariant distribution of the process where, in each period, agents choose to maximize $\ln\left(\pi_i^N(i) v_i + \sum_{j \in N \setminus i}\pi_{i}^N(j) \hat{p}_j^N(x)\right) + \epsilon_x$ where $\hat{p}_j^N(x)$ is the empirical frequency thus far of the choice of $x$ by agent $j$ in group N.} Under these two assumptions, the long run choice frequencies are given by Equation \ref{EQ:linInMeans}.

Our third interpretation concerns multidimensional performance in team environments, such as sports teams, organizations, firms, and governments.\footnote{We thank an anonymous referee for suggesting this additional interpretation.} We find this interpretation particularly compelling when the assumption of positive $\pi_i^N(j)$ is dropped, which we consider in Section \ref{sec:negweights}. For example, in a basketball setting, let $v_i$ denote player $i$’s intrinsic multidimensional performance profile, including dimensions such as points, assists, rebounds, steals, blocks, or turnovers. When player $i$ participates in a lineup $N$, the observed performance vector $p_i^N$ is not simply $v_i$, because performance is shaped by the roles and contributions of the other players in the lineup. Our model captures this interaction. The coefficient $\pi_i^N(i)$ measures the extent to which player $i$’s observed performance reflects their own intrinsic profile, while $\pi_i^N(j)$ measures the influence of teammate $j$’s observed performance on player $i$’s performance in lineup $N$. 

Thus, a player who is intrinsically a scorer may take fewer shots when paired with another high-usage scorer, while increasing assists or off-ball contributions. Likewise, a rebounder’s observed rebounding may depend on whether the lineup includes other strong rebounders. Since players appear in multiple overlapping lineups, variation in team composition generates identifying variation. The multidimensionality of the outcome is crucial: changes in points alone may be ambiguous, whereas the full vector of points, assists, rebounds, and defensive actions reveals how roles are reallocated across lineups. This makes it possible to distinguish intrinsic performance profiles from peer or lineup-composition effects.

\section{Results}\label{Sec:Results}

In this section, we study the behavioral implications of the general, the Luce, and the uniform linear-in-means models. We begin with the general case and provide two characterizations of datasets consistent with the general case as well as partial and point identification results for $v_i$ and $\pi^N$. Our first characterization is via the non-empty intersection of a collection of convex sets. These convex sets correspond to the set of $v_i$ which feasibly induce the observed choice of agent $i$ the context of a given group. Our partial identification of $v_i$ builds on this result. Our second characterization provides an existential linear program which fails to hold if and only if the dataset is consistent with the general linear-in-means model. This linear program can be interpreted as an across-group consistency condition about how an agent's behavior differentiates them from the behavior of other agents. We then proceed to the Luce and uniform linear-in-mean models and provide refinements of these results.

\subsection{The General Model}\label{Sec:GLM}
We now begin our analysis of the general linear-in-means model. Recall that in GLM, for each group of agents $N$, each agent's choice $p_i^N$ is written as a convex combination of $v_i$ and $p_j^N$ across all $j \in N \setminus i$. Further, there are no restrictions across groups about the weights assigned to each agent other than each agent $i$ puts some weight on $v_i$. This means that testing GLM amounts to finding if there is some $v_i$ which can induce, for all $N \in \mathcal{N}_i$, $p_i^N$ as a convex combination of $v_i$ and $p_j^N$. This observation tells us two things. First, testing GLM can be done agent by agent. Whether or not agent $i$ has a feasible rationalizing $v_i$ can be tested independently of agent $j$. Second, a key part of testing GLM is finding the set of feasible $v_i$ for agent $i$ given $p^N$ for each $N \in \mathcal{N}_i$.

\begin{figure}[h]
\begin{center}
 \includegraphics[width=7cm]{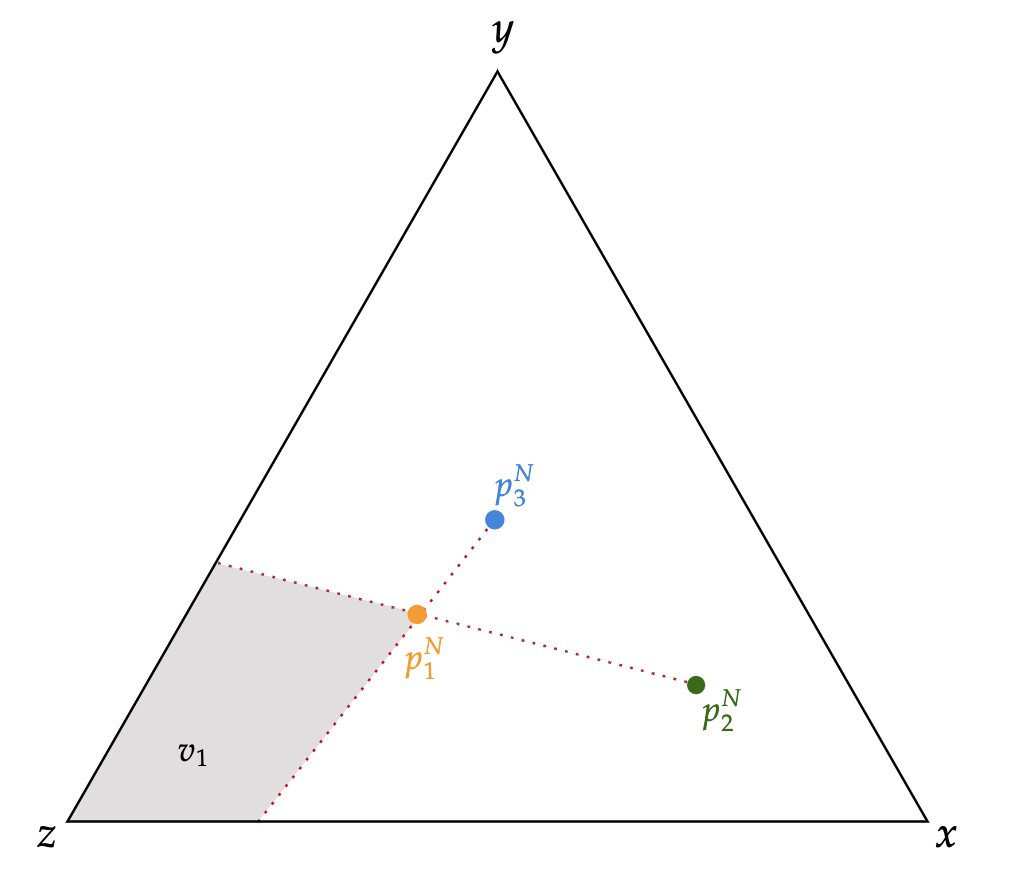}   
\end{center}
\caption{\small{The figure considers three agents choosing when three alternatives are available. In the general linear-in-means model, $p_1^N$ is an arbitrary convex combination of $v_1$, $p_2^N$, and $p_3^N$. The shaded region corresponds to the set of feasible $v_1$ which could induce $p_1^N$ given $p_2^N$ and $p_3^N$. This corresponds to $co^{-1}(\Delta(p^N),p_1^N)$ from Equation \ref{EQ:InverseCone}.} }
\label{Fig:FeasibleCone}
\end{figure}

With this in mind, suppose we observe data $p^N$ and we see that agent $i$'s choices, $p_i^N$, lie on the relative interior of the convex hull of the other agents' choices, which we write $int \Delta(p_{-i}^N)$. In this case, no matter what $v_i \in \Delta(X)$ we consider, $p_i^N$ can be written as a convex combination of the other agents' choices. Thus we can simply ask that agent $i$ puts a vanishingly small amount of weight on $v_i$ in their convex combination and rationalize this $v_i$. Now suppose we observe data such that $p_i^N$ does not lie in $int \Delta(p^N_{-i})$. In this case, a $v_i$ is feasible if and only if $p_i^N$ can be written as a convex combination of $v_i$ and some convex combination of $p_j^N$ for $j \in N \setminus i$. Thus the set of points on the ``opposite side" of $p_i^N$ from $\Delta(p^N_{-i})$ correspond to the set of feasible $v_i$. Formally, this is given by the following equation.
\begin{equation}\label{EQ:InverseCone}
    co^{-1}(\Delta(p^N),p_i^N)=\{v\in \Delta(X)|v = \sum_{j \in N}\gamma_j p_j^N, \gamma_j \leq 0 \text{ } \forall j \in N\setminus i, \sum_{j \in N}\gamma_j = 1\}
\end{equation}
Figure \ref{Fig:FeasibleCone} visualizes the set of feasible $v_i$ for agent $1$ when there are three goods and three agents. These two observations tell us that the testable content of GLM amounts to checking, for each $i$, if there is some feasible $v_i$ which is in each $co^{-1}(\Delta(p^N),p_i^N)$.

Our previous observations tell us that testing GLM amounts to checking if, for each $i$, a collection of polyhedral sets has a point of mutual intersection.\footnote{A related class of problems was studied by \citet{samet1998common} and \citet{morris1994trade}. These papers examine when agents share a common prior and characterize this property through a no-money-pump condition. The condition we introduce below is analogous in spirit, but admits a direct interpretation in terms of peer influence.} Alternatively, testing this mutual intersection condition can be framed as a linear program. To this end, we consider vectors $b^N$ which each correspond to directions by which an agent can be behaviorally differentiated within group $N$. Our condition requires a minimal amount of consistency across groups in the way that an analyst can behaviorally differentiate a given agent.

\begin{definition}
     We say that a set of vectors $\{b^N\}_{N \in \mathcal{N}_i}$ with $b^N \in \mathbb{R}^X$ for each $N \in \mathcal{N}_i$, \textbf{behaviorally differentiates} agent i if the following are satisfied.
     \begin{enumerate}
         \item (Alignment) For each $N \in \mathcal{N}_i$, $b^N \cdot p_i^N > 0$.
         \item (Differentiation) For each $N \in \mathcal{N}_i$ and each $j \in N \setminus i$, $b^N\cdot (p_i^N - p_j^N) \geq 0$.
     \end{enumerate}
\end{definition}

Our interpretation of the vector $b^N$ as behaviorally differentiating agent $i$ relies on the fact that we impose that $b^N$ both aligns with agent $i$'s choice in group $N$ and the fact that $b^N$ is weakly more aligned with agent $i$'s choice in group $N$ than any other agent's choice in group $N$. Specifically, alignment asks that $b^N$ and the behavior of agent $i$ in group $N$ both agree, approximately, on their direction. Furthermore, differentiation asks that $b^N$ agrees with the behavior of agent $i$ in group $N$ (weakly) more than the behavior of any other agent in group $N$.

Our focus on $\{b^N\}_{N\in \mathcal{N}}$ which satisfy alignment and differentiation is due to their ability to reveal information about the intrinsic portion of agent $i$'s utility. In our case, this corresponds to agent $i$'s ideal point $v_i$. In the linear-in-means model, an agent $i$'s choice is a convex combination of their own ideal point and the choices of the other agents in their group. Thus, if an analyst can find $\{b^N\}_{N\in \mathcal{N}}$ which both align and strictly differentiates agent $i$, each $b^N$ must be pointing in the direction of agent $i$'s ideal point, approximately. Hence, the vectors in $\{b^N\}_{N\in \mathcal{N}}$ should exhibit a corresponding form of cross-group consistency. This leads us to our next definition.

\begin{definition}
    We say that a dataset $\{p^N\}_{N \in \mathcal{N}}$ \textbf{consistently differentiates} agent $i$ if every $\{b^N\}_{N\in \mathcal{N}_i}$ which behaviorally differentiates agent $i$ does not satisfy $\sum_{N \in \mathcal{N}_i}b^N \ll 0$.\footnote{When working with vectors, we use $b \ll c$ to denote that the vector $b$ is strictly less than $c$ in each of its dimensions. We use $b < c$ to denote that vector $b$ is weakly less than $c$ in each of its dimensions and strictly less in at least one dimension. We use $b \leq c$ to denote that vector $b$ is weakly less than $c$ in each of its dimensions.}
\end{definition}

This definition asks that, no matter how the analyst chooses the behaviorally differentiating vectors $\{b^N\}_{N\in \mathcal{N}}$, there is no way for the vectors to cancel out across groups while also aligning with agent $i$'s choices, on average.\footnote{Letting $b = \sum_{N \in \mathcal{N}_i}b^N \ll 0$, we know that $b \cdot p_i^N < 0$ since $b$ is a strictly negative vector and $p_i^N$ is a weakly positive non-zero vector. This corresponds to $\{b^N\}_{N\in \mathcal{N}}$, on average, misaligning with agent $i$'s choices.} Thus, in the case of the linear-in-means model, failures of consistency tell us that an agent's bliss point must vary with their group. We are now ready to state our characterization of GLM.

\begin{theorem}\label{Thm:GenChar}
For a dataset $\{p^N\}_{N \in \mathcal{N}}$, the following are equivalent.
\begin{enumerate}
    \item $\{p^N\}_{N \in \mathcal{N}}$ is consistent with the general linear-in-means model.
    \item For every $i \in \mathcal{A}$, the collection of sets $\{co^{-1}(\Delta(p^N),p_i^N)\}_{N \in \mathcal{N}_i}$ has a point of mutual intersection.
    \item For all agents $i \in \mathcal{A}$, $\{p^N\}_{N \in \mathcal{N}}$ consistently differentiates agent $i$.
\end{enumerate}
\end{theorem}

We leave all proofs to the appendix. The equivalence between (1) and (2) follows from our discussion at the start of Section \ref{Sec:GLM}. If $co^{-1}(\Delta(p^N),p_i^N)$ corresponds to the set of feasible $v_i$ for agent $i$ in group $N$, then there needs to be some $v_i$ common to this set across all groups. The equivalence between (1) and (3) is partially a result of (2). As mentioned previously, since testing for GLM amounts to testing for a point of mutual intersection, we can transform this via linear programming duality to get our consistent differentiation condition. However, we note that our condition does not follow immediately from (2) and an application of the result of \citet{samet1998common} and \citet{billot} as, in our proof, we work with polyhedral sets rather than compact sets. This variation allows us to recover the exact form of our condition.\footnote{In Appendix \ref{app:Samet} we discuss the relation and application of \citet{samet1998common} to our Theorem \ref{Thm:GenChar}.}

Before moving on, we note that condition (2) from Theorem \ref{Thm:GenChar} reduces to an easily checkable condition in the one-dimensional case. In the one-dimensional case, the choice frequency of one alternative is a sufficient statistic for the choice frequency of the other alternative. To simplify notation in the one dimension case, we use $p_i^N$ to denote the probability that agent $i$ chooses alternative $x$ in group $N$. Now let $\mathcal{N}_i^- \subseteq \mathcal{N}_i$ denote the set of groups $N$ satisfying $p_i^N \leq p_j^N$ for each $j \in N \setminus i$. Similarly, let $\mathcal{N}_i^+ \subseteq \mathcal{N}_i$ denote the set of groups $N$ satisfying $p_i^N \geq p_j^N$ for each $j \in N \setminus i$.

\begin{corollary}\label{cor:1DimChar}
    In the one dimension case, a dataset $\{p^N\}_{N \in \mathcal{N}}$ is consistent with the general linear-in-means model if and only if, for all $i\in\mathcal{A}$, the following conditions hold.
    \begin{enumerate}
        \item $\min_{N \in \mathcal{N}_i^-} p_i^N \geq \max_{\mathcal{N}_i^+} p_i^N$ when both $\mathcal{N}_i^-$ and $\mathcal{N}_i^+$ are non-empty,
        \item $\{p_i^N\}_{N \in \mathcal{N}_i^- \cap \mathcal{N}_i^+}$ contains at most one value (which we denote $p_i^=$),
        \item $\min_{N \in \mathcal{N}_i^-} p_i^N = p_i^= = \max_{\mathcal{N}_i^+} p_i^N$ when $\mathcal{N}_i^- \cap \mathcal{N}_i^+$ is non-empty.
    \end{enumerate}
\end{corollary}

\subsubsection{Identification}\label{Sec:Ident}

We now turn our attention to the identification properties of GLM. Specifically, our focus is on the joint recovery of an agents' bliss point as well as the network structure of each group. Going back to \citet{manski1993identification}, it is known that recovering influence parameters is generally a hard problem. In this section we highlight that, conditional on observing group variation, many of these identification problems are due to the one-dimensional formulation of the original reflection problem. Our identification argument for $(v_i,\pi_i^N)$ proceeds in two steps. First, we show that, generically and with specific group variation, we are able to recover $v_i$ so long as $|X| \geq 3$ (i.e. in any case other than the one dimension case). Further, this identification generically fails in the one dimension case. Second, conditional on recovering $v_i$, we show that $\pi_i^N(j)$ is generically identified from $v_i$ and $p_{-i}^N$ so long as $|X| \geq |N|$. These two results highlight that multiple dimensions are useful in recovering each agent's bliss point, $v_i$, as well as their peer influence parameters, $\pi_i^N$. 

\subsubsection*{Recovery of Ideal Points} We begin by discussing the partial identification of $v_i$ that arises without sufficient group variation.

\begin{definition}
    For parameter $v_i$, we say that $A$ is the \textbf{sharp identified set} if $v_i \in A$ if and only if there exist $\pi_i^N$ for each $N \in \mathcal{N}_i$ such that $p_i^N=\pi_i^N(i) v_i + \sum_{j \in N \setminus i}\pi_{i}^N(j) p_j^N$. Further, we say that $v_i$ is \textbf{point identified} if the partially identified set $A$ is singleton.
\end{definition}

Our first goal is to characterize the sharp identified set of $v_i$. Recall our discussion of $co^{-1}(\Delta(p^N),p_i^N)$ prior to Theorem \ref{Thm:GenChar}; it corresponds to the set of feasible $v_i$ given observed choices $p^N$. Theorem \ref{Thm:GenChar} tells us that our dataset is consistent with GLM if and only if, for each $i \in \mathcal{A}$, $\{co^{-1}(\Delta(p^N),p_i^N)\}_{N \in \mathcal{N}_i}$ have a point of mutual intersection. This amounts to testing if there is some ideal point $v_i$ which is feasible in each group $N \in \mathcal{N}_i$. It follows from similar logic that any point in the mutual intersection of $\{co^{-1}(\Delta(p^N),p_i^N)\}_{N \in \mathcal{N}_i}$ is a feasible value for $v_i$.

\begin{proposition}\label{Prop:genidentification}
    In GLM, the sharp identified set for $v_i$ is given by $\bigcap_{N \in \mathcal{N}_i} co^{-1}(\Delta(p^N),p_i^N)$.
\end{proposition}

Proposition \ref{Prop:genidentification} formally states the observation from the previous paragraph. As mentioned earlier, one of our main goals is to give conditions under which $v_i$ and $\pi_i^N$ are jointly point identified. Here, point identified means that the sharp identified set is a singleton. As a first step in our two step procedure, our next result gives a sufficient condition for point identification of $v_i$. Let $\mathcal{N}_i^{ext} \subseteq \mathcal{N}_i$ denote the set of groups $N$ with $p_i^N \not \in \Delta(p_{-i}^N)$.

\begin{corollary}\label{cor:PointIdent}
    Let $N_j=\{i,j\}$ and $N_k=\{i,k\}$. Suppose that $\{p^N\}_{N \in \mathcal{N}}$ is consistent with GLM, $N_j,N_k \in \mathcal{N}_i^{ext}$, and that the vectors $(p_i^{N_j}-p_j^{N_j})$ and $(p_i^{N_k}-p_k^{N_k})$ are linearly independent. Then $v_i$ is point identified.
\end{corollary}

\begin{figure}[h!]
\begin{center}
 \includegraphics[width=7cm]{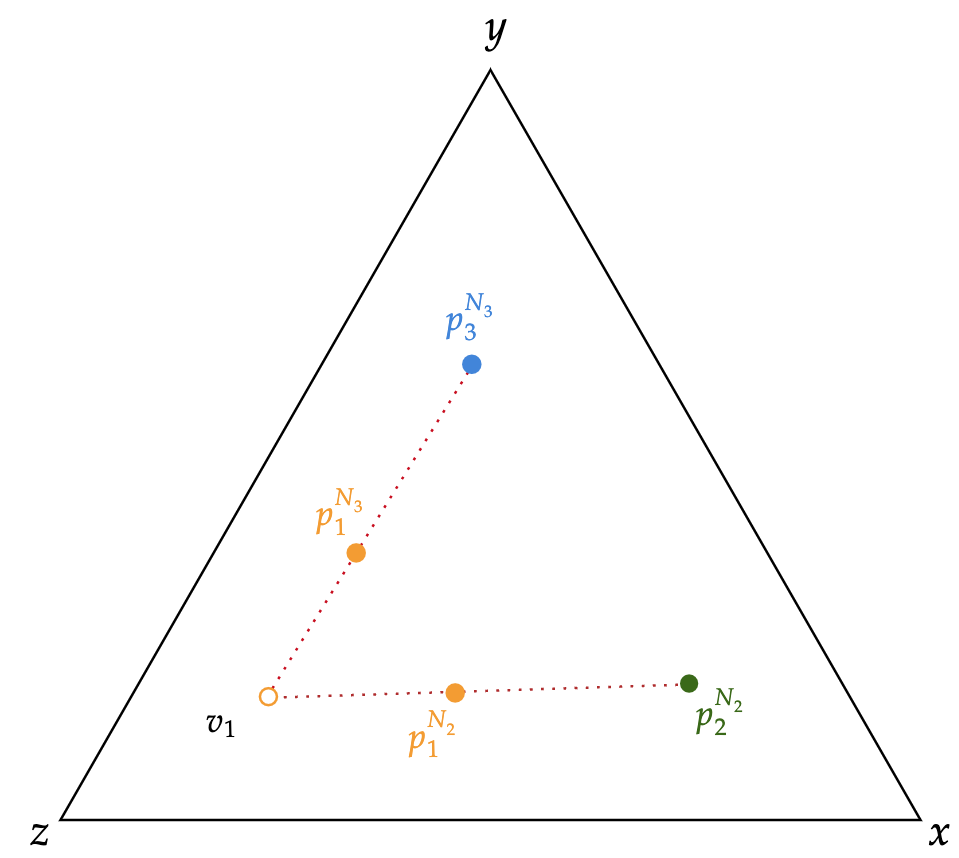}   
\end{center}
\caption{\small{The figure considers choice by agents $1$, $2$, and $3$. We observe choice in two groups; $N_2=\{1,2\}$ and $N_3=\{1,3\}$. Consider the two rays which trace out $co^{-1}(\Delta(p^N),p_1^N)$ for each group. These rays intersect, so, by Theorem \ref{Thm:GenChar}, this dataset is consistent with GLM. The linear independence of $(p_1^{N_2}-p_2^{N_2})$ and $(p_1^{N_3}-p_3^{N_3})$ gives us that these rays intersect at a single point. This single point corresponds to the uniquely feasible $v_1$. } }
\label{Fig:PointIdent}
\end{figure}

Figure \ref{Fig:PointIdent} offers a visualization of Corollary \ref{cor:PointIdent}. %While we do not formally show it here, Corollary \ref{cor:PointIdent} naturally extends. Consider $co^{-1}(\Delta(p^N),p_i^N)$ for a group of agents $N$. Note that, in the case of $N \in \mathcal{N}_i^{ext}$, if we drop the non-negativity restriction on $v$ in our definition of $co^{-1}(\Delta(p^N),p_i^N)$, then $co^{-1}(\Delta(p^N),p_i^N)$ defines a polyhedral cone. The extremal rays of this cone take the form $p_i^N-p_j^N$ plus some location translation. In out setting, linear independence of the extremal rays of two of these $n$ dimensional cones will make the intersection of these two sets $(n-1)$ dimensions. Thus, when we observe $n$ groups of $n$ agents, with each group being contained in $\mathcal{N}_i^{ext}$, and the set of extremal rays of $co^{-1}(\Delta(p^N),p_i^N)$ across all $n$ groups of $n$ agents being linearly independent, $\bigcap_{N \in \mathcal{N}_i^{ext}}\{co^{-1}(\Delta(p^N),p_i^N)\}$ is a $(n-n)=0$ dimensional set. By Theorem \ref{Thm:GenChar} this set is non-empty and thus we get point identification. 
We now return to our example with Ann, Ben, and Can to highlight Corollary \ref{cor:PointIdent}.

\begin{example}\label{Ex:IdealPointIdent}
    Recall the choices of Ann, Ben, and Can. While Can is studying abroad, Ann chooses $(0.5,0.1,0.4)$, (tennis, volleyball, walking), and Ben chooses $(0.7,0.1,0.2)$. While Ben is studying abroad, Ann chooses $(0.1,0.5,0.4)$ and Can chooses $(0.1,0.7,0.2)$. We now calculate the difference between Ann's choices and Ben' and Can's choices in their respective groups. When both Ann and Ben are choosing, the difference in their choices is given by $(0.2,0,-0.2)$. Similarly, when both Ann and Can are choosing, the difference between their choices is given by $(0,0.2,-0.2)$. As these two vectors are linearly independent, Corollary \ref{cor:PointIdent} tells us that we can point identify Ann's ideal point. Ann's ideal point is given by the point of intersection between the two lines $(0.5,0.1,0.4)+t(0.2,0,-0.2)$ and $(0.1,0.5,0.4)+r(0,0.2,-0.2)$. The intersection of these two lines is guaranteed by Theorem \ref{Thm:GenChar}. Calculating the point of intersection gives us $(0.1,0.1,0.8)$, thus identifying Ann's ideal point. This also tells us that, even though walking is never Ann's most chosen activity, it is her most preferred activity in isolation.
\end{example}

Corollary \ref{cor:PointIdent} gives simple intuition as to how one can recover $v_i$ using simple data, two groups of two agents. It may often be the case that an analyst does not have access to data from groups of this specific form. With this in mind, we now extend the point identification results of Corollary \ref{cor:PointIdent} to arbitrary collections of arbitrary groups.

\begin{theorem}\label{thm:pointIdent}
    Fix agent $i$ and suppose that $\{p^N\}_{N \in \mathcal{N}}$ is consistent with GLM. If, for each non-zero vector $p$ in the span of $\{p_i^N-p_j^N\}_{N \in \mathcal{N}_i,j \in N\setminus i}$, there is some group $M \in \mathcal{N}_i$ such that $p$ is not in the span of $\{p_i^M-p_j^M\}_{j \in M \setminus  i}$, then $v_i$ is point identified.
\end{theorem}

The condition used in Theorem \ref{thm:pointIdent} is effectively a linear independence condition. It works by first taking, for each $N \in \mathcal{N}_i$, the set $co^{-1}(\Delta(p^N),p_i^N)$ and embedding it in the minimal affine subspace containing it and then imposing that these affine subspaces have a single point of mutual intersection. Since the data is assumed to be consistent with GLM, this single point of mutual intersection is also the single point in $\bigcap_{N \in \mathcal{N}_i} co^{-1}(\Delta(p^N),p_i^N)$, thus giving us point identification of $v_i$. While the condition in Theorem \ref{thm:pointIdent} is stated as checking a condition for an infinite collection of vectors, we note that Theorem \ref{thm:pointIdent} can be checked in finite time. Instead of checking every vector in the span of $\{p_i^N-p_j^N\}_{N \in \mathcal{N}_i,j \in N\setminus i}$, one could form a basis for this linear subspace and checking the condition for each element of the basis would be sufficient. Further, unlike in Corollary \ref{cor:PointIdent} where we could specify that an analyst needs exactly two groups of two agents for point identification, in Theorem \ref{thm:pointIdent}, the number of groups necessary for identification is heavily dependent on the realized data. Even for groups of arbitrary size, it is possible that the conditions of Theorem \ref{thm:pointIdent} could be satisfied in the case where the analyst only observes two groups. On the other hand, even if each additional group reduces the dimension of the partially identified set $co^{-1}(\Delta(p^N),p_i^N)$, it may take up to $|N_{min}^i|$ groups to recover point identification, where $N_{min}^i$ is the smallest group in $\mathcal{N}_i$.

\subsubsection*{Failures of Identification with One-Dimension}
\begin{figure}[h!]
    \centering
    \includegraphics[width=0.5\linewidth]{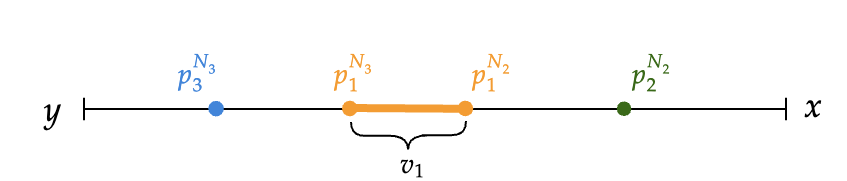}
    \caption{\small{The figure considers choice by agents $1$, $2$, and $3$. We observe choice in two groups; $N_2=\{1,2\}$ and $N_3=\{1,3\}$. Unlike in Figure \ref{Fig:PointIdent}, $v_1$ is not point identified here. This is due to the fact that, in the one dimension case, $(p_1^{N_2}-p_2^{N_2})$ and $(p_1^{N_3}-p_3^{N_3})$ will always be linearly dependent; they will be some multiple of $(1,-1)$. Geometrically, the rays corresponding to $co^{-1}(\Delta(p^N),p_1^N)$ for $N_2$ and $N_3$ can intersect at a single point if and only if $p_1^{N_2}=p_1^{N_3}$.}}
    \label{fig:1DNoIdent}
\end{figure}

Before moving on, we point out that point identification of $v_i$ generically fails in the one-dimensional case (two-outcome case). Recall the three conditions of Corollary \ref{cor:1DimChar}. The last two conditions arise as a result of non-generic situations. Specifically, if we observe data where $p_i^N \neq p_i^M$ for all $N,M \in \mathcal{N}_i$, then it is sufficient to just check the first condition. However, in this case, $v_i$ is necessarily not point identified. In the one-dimensional case, the sharp identified set corresponds to the interval $[\max_{\mathcal{N}_i^+} p_i^N, \min_{N \in \mathcal{N}_i^-} p_i^N]$ and this set is singleton only in a non-generic case. We visualize this failure of point identification in Figure \ref{fig:1DNoIdent} and show it in Example \ref{Ex:FailPointIdent}.

\begin{example}\label{Ex:FailPointIdent}
    Consider a counterfactual world with Ann, Ben, and Can where their college no longer has tennis courts. In this world, Ann, Ben, and Can have the choice of splitting their exercise time between volleyball and walking. While Can is studying abroad, Ann's choices are given by $(0.8,0.2)$, (volleyball, walking), and Ben's choices are given by $(0.9,0.1)$. While Ben is studying abroad, Ann's choices are given by $(0.7,0.3)$ and Can's choices are given by $(0.85,0.15)$. When both Ann and Ben are choosing, the difference in their choices is given by $(-0.1,0.1)$. On the other hand, when both Ann and Can are choosing, the difference in their choices is given by $(-0.15,0.15)$. Notice that these two vectors are multiples of each other and are thus linearly dependent. These choices, as would any other choices in this two-alternative world, fail the conditions of our Corollary \ref{cor:PointIdent}. Ann's choices when Ben is abroad are given by $(0.7,0.3)$ and, for that to lie in the convex hull of $(\alpha,1-\alpha)$ and $(0.85,0.15)$, $\alpha$ must be weakly less than $0.7$. Similarly, her choices when Can is abroad are given by $(0.8,0.2)$ and, for this to be between $(0.9,0.1)$ and $(\alpha,1-\alpha)$, $\alpha$ must be weakly less than $0.8$. Combining our two constraints gives us that any vector $(\alpha,1-\alpha)$ which satisfies $0 \leq \alpha \leq 0.7$ is a potential ideal point for Ann, hence we cannot point identify Ann's ideal point.
\end{example}

\subsubsection*{Recovery of Network Effects}

We now turn our attention to the recovery of $\pi_i^N$. This recovery utilizes knowledge of $v_i$ and thus is dependent on our previous arguments in Theorem \ref{thm:pointIdent} and Corollary \ref{cor:PointIdent}.

\begin{proposition}\label{Prop:PiPointIdent}
    Suppose that $v_i$ is point identified. Then $\pi_i^N$ is point identified if the set of vectors including $v_i$ and $\{p_j^N\}_{j \in N\setminus i}$ is affinely independent.\footnote{A set of points $\{p_j\}$ is affinely independent if for each $i$, there is no collection $\{\lambda_j\}_{j\neq i}$ for which $\sum_{j\neq i}\lambda_j =1$ where $p_i = \sum_{j\neq i}\lambda_j p_j$. When each point $p_j$ satisfies $p_j \geq 0$ and $\sum_{x \in X}p_j(x)=1$, affine independence and linear independence are equivalent.}
\end{proposition}

The identification of $\pi_i^N$ follows from basic properties of simplices. Specifically, any point in the convex hull of a simplex can be written as a convex combination, with unique weights, of the extreme points of the simplex. The affine independence condition ensures that $v_i$ and $\{p_j^N\}_{j \in N\setminus i}$ form the extreme points of a simplex. We point out that affine independence of $v_i$ and $\{p_j^N\}_{j \in N\setminus i}$ puts joint restrictions on the number of agents and alternatives. Notably, when $|N| \leq |X|$, the set of points given by $v_i$ and $\{p_j^N\}_{j \in N\setminus i}$ is generically affinely independent. On the other hand, if $|N| > |X|$, the set of points $v_i$ and $\{p_j^N\}_{j \in N\setminus i}$ is never affinely independent. In terms of dimension counting, in group $N$, there are $|N|$ agents each with $|N|-1$ unknowns giving a total of $|N|(|N|-1)$ unknowns in $\pi^N$. Each agent has $|X|-1$ observables per group, and so the total number of observables in group $N$ is given by $|N|(|X|-1)$. This tells us that we have more observables than unknowns if $|X| \geq |N|$. In order to highlight this result, we return to Ann, Ben, and Can.

\begin{example}\label{Ex:WeightPointIdent}
    In Example \ref{Ex:IdealPointIdent}, we were able to identify Ann's ideal point as $(0.1,0.1,0.8)$. During the semester when Can is abroad, Ben's choices correspond to $(0.7,0.1,0.2)$. Since Ann's ideal point and Ben's choices are distinct points, they are trivially affinely independent. By Proposition \ref{Prop:PiPointIdent}, this means that we are able to recover both the impact of Ben's choices on Ann's choices and the impact of Ann's own ideal point on her choices. This problem amounts to solving for $\pi_{Ann}$ in the equation $(0.5,0.1,0.4)=\pi_{Ann}(0.1,0.1,0.8) + (1-\pi_{Ann})(0.7,0.1,0.2)$. Doing so gives us that $\pi_{Ann}$ is equal to $\frac{1}{3}$. Thus the relative importance of Ann's ideal point is $\frac{1}{3}$ and the relative importance of Ben's choices is $\frac{2}{3}$. Due to the symmetry of the problem, these weights are the same when considering the choices of Ann and Can.
\end{example}

Our results in this section show how utilizing multidimensional outcome variables, along with group variation, allows the analyst to jointly recover agents' bliss points and the underlying network structure. We take this to be prescriptive. While outcome variables can sometimes be sufficiently summarized in a one-dimensional statistic, such as labor time, by gathering more granular data and considering a higher dimensional outcome variable, we can actually recover and quantify heterogeneous peer effects (subject to the conditions of our results).

% {\color{red} EDIT TO SAY WE TALK ABOUT MANSKI LATER IN SECTION 4.2 OR JUST GET RID OF IT AND LEAVE 4.2 We conclude our discussion of identification by putting our results in the context of the reflection problem of \citet{manski1993identification}.  Our focus is on a related but slightly different version of the linear-in-means model proposed in \citet{manski1993identification}. First, the primary specification of our of the model is primarily concerned with conformism ($\pi_i^N \geq 0$ and $\sum_{j \in N}\pi_i^N(j)=1$). Second, we allow for weighted averages (from unknown networks) while the version of \citet{manski1993identification} assumes an unweighted average across peers. We make weaker assumptions in one direction and stronger assumptions in other directions. \citet{manski1993identification} shows that identification fails in his setup. This continues to be true even if our conformism assumptions are applied to the setup of \citet{manski1993identification}, primarily due to the fact that the reflection problem considers one-dimensional outcome variables.}

\subsection{Luce Linear-in-Means}

As we have seen, there are conditions in GLM under which we are able to recover both an agent's ideal point $v_i$ and their social interaction parameters $\pi_i^N$ for groups that we observe. However, these identified parameters have little predictive power for groups we do not observe in GLM. Since GLM allows arbitrary variation in $\pi_i^N$ across groups, the most we can predict for agent $i$'s choice in an unobserved group $N$ is that it lies within the convex hull of $\{v_j\}_{j \in N}$. Our goal in this section is to consider the Luce linear-in-means model which allows us to connect our social interaction parameters across groups. LLM allows analysts to predict choices in unobserved groups subject to the conditions for identification from Section \ref{Sec:Ident} holding.

Recall that LLM supposes that each agent $i$ has a weighting function, $w_i(j)$, which corresponds to the absolute importance of agent $j$ to agent $i$. In a group $N$, the relative importance of agent $j$ to agent $i$ is given by the renormalization of this weighting function, $\pi_i^N(j)=\frac{w_i(j)}{\sum_{k \in N} w_i(k)}$. The interpretation here is that there is a single underlying network fixed across each group. We first discuss the identification and predictive properties of LLM. We focus on the case when $v_i$ is known. For a specific group $N$, if the conditions of Proposition \ref{Prop:PiPointIdent} hold, we can pin down $\pi_i^N$. This allows us to pin down the relative weights of each agent $j \in N\setminus i$ as $\frac{w_i(j)}{w_i(k)}= \frac{\pi_i^N(j)}{\pi_i^N(k)}$. This is the exact condition for identification that is used in the Luce model of \citet{luce1959individual}. This means that if $N = \mathcal{A}$, we can predict choice (for agent $i$) on every possible group of agents. However, we may not always observe choice on $N = \mathcal{A}$ or, if we do, the conditions of Proposition \ref{Prop:PiPointIdent} may not be satisfied at $N=\mathcal{A}$. To regain full predictive power in LLM, we need the conditions of Proposition \ref{Prop:PiPointIdent} to hold for a collection of groups $\{N_l\}_{l=1}^L$ such that every two agents $k$ and $j$ can be compared through a string of groups.

\begin{definition}
    Suppose $v_i$ is point identified. Agents $j$ and $k$ are \textbf{comparable by agent i} if there exist two observed groups $N_1, N_2 \in \mathcal{N}$ such that the following hold.\begin{itemize}
        \item $i \in N_1 \cap N_2$,
        \item $j \in N_1$ and $k \in N_2$,
        \item For each $l \in \{1,2\}$, $v_i$ and $\{p_m^{N_l}\}_{m \in N_l \setminus i}$ are affinely independent.
    \end{itemize}
\end{definition}

\begin{proposition}\label{Prop:LuceIdent}
    Suppose that $v_i$ is point identified. Then $w_i(j)$ is point identified for all $j \in \mathcal{A}$ if every pair of agents $j \neq k$ are comparable by agent $i$.
\end{proposition}

Proposition \ref{Prop:LuceIdent} is an immediate result of our Proposition \ref{Prop:PiPointIdent}, Corollary 4 of \citet{alos2024characterization}, and the fact that $i$ is in both $N_1$ and $N_2$. Further, Proposition \ref{Prop:LuceIdent} gives conditions under which we can predict the choice of agent $i$ in any group $N$, conditional on knowing $v_j$ for each $j \in N$.  We show how to operationalize this predictive power by revisiting Ann, Ben, and Can one last time.

\begin{example}\label{Ex:LLMPrediction}
    Recall the choices of Ann, Ben, and Can described in Example \ref{Ex:IdealPointIdent}. Our goal now is to predict how Ann, Ben, and Can will choose when all three of them are choosing together. As we showed in Example \ref{Ex:WeightPointIdent}, the weight Ann puts on Ben or Can is twice the weight she puts on her own ideal point. This tells us that Ann's $w$ vector is given by $(0.2,0.4,0.4)$. Since we have only observed Ben and Can choosing in a single group each, we are unable to pin down their ideal points or network weights. Now suppose we observe choice from a third semester when Ann chooses to study abroad. In this semester, Ben and Can choose $(\frac{19}{30},\frac{11}{30},0)$ and $(\frac{11}{30},\frac{19}{30},0)$ respectively. Since neither Ben nor Can decide to walk as a form of exercise this semester, it reveals that both of their ideal points put no weight on walking. Using this observation and similar techniques to those in Example \ref{Ex:IdealPointIdent}, we can conclude that Ben's and Can's ideal points are $(0.9,0.1,0)$ and $(0.1,0.9,0)$ respectively. By applying Proposition \ref{Prop:PiPointIdent} and \ref{Prop:LuceIdent}, we recover that Ben and Can put the same amount of weight on their own ideal points and the choices of each other agent. This means that both of their $w$ vectors are $(\frac{1}{3},\frac{1}{3},\frac{1}{3})$. Since we have recovered the ideal points and the $w$ vectors of Ann, Ben, and Can, we can now predict their choices in the semester where all three of them are choosing together. In this semester, Ann's choices are $(\frac{11}{30},\frac{11}{30},\frac{4}{15})$, Ben's choices are $(\frac{8}{15},\frac{1}{3},\frac{2}{15})$, and Can's choices are $(\frac{1}{3},\frac{8}{15},\frac{2}{15})$.
\end{example}

\begin{figure}[h]
\centering
    \includegraphics[width=.4\linewidth]{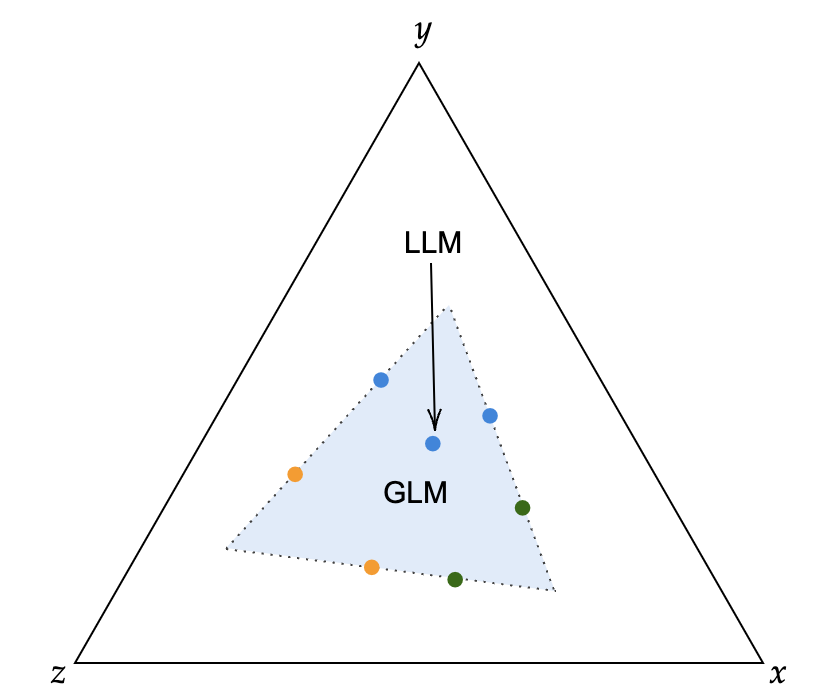}

\caption{\small{This figure illustrates predictions of GLM and LLM for three alternatives, $X = \{x, y, z\}$ and three agents: blue, green, and orange. The blue-shaded triangle identifies all possibilities for the choice probability of the blue agent consistent with GLM in the entire group, given three binary group choices for each agent depicted on three edges of the shaded region. Note that LLM predicts uniquely the choice probability of the blue agent.}}
\label{fig:DifferentModels}
\end{figure}

Figure \ref{fig:DifferentModels} compares GLM and LLM in terms of their prediction power on the Marschak-Machina triangle in a domain of three alternatives, $X = \{x, y, z\}$ and three agents $\{1,2,3\}$. This is the simplest environment in which we can illustrate the differences between these models. In the figure, stochastic choices for three binary groups ($\{1,2\}, \{1,3\}$ and $\{2,3\}$) are fixed and denoted by different colored dots on the edges of the dotted triangle. The blue dots indicate agent $3$'s choices, $p_3^{\{1,3\}}$ and $p_3^{\{2,3\}}$.  We then illustrate the predictions for $p_3^{\{1,2,3\}}$ for GLM and LLM. The blue-shaded triangle identifies all possibilities for $p_3^{\{1,2,3\}}$ for GLM given these binary group choices. This figure illustrates both the predictive and explanatory power of GLM: If $p_3^{\{1,2,3\}}$ lies outside this triangle, the data cannot be explained by GLM.

LLM restricts choices for $p_3^{\{1,2,3\}}$ even further. Indeed, LLM predicts that there is only a single possibility for $p_3^{\{1,2,3\}}$ indicated by a blue dot inside GLM's prediction. If  $p_3^{\{1,2,3\}}$ is not equal to this point, then the data cannot be explained by the Luce linear-in-means model. Note that the prediction of LLM belongs to the shaded triangle indicating that LLM is a special case of GLM.

While we focus on LLM to recover predictive power in the linear-in-means model, one could instead consider any known mapping from $\pi_i^N$ to $\pi_i^M$ for two different groups $N$ and $M$ to recover predictive power. We focus on LLM for two reasons. First, LLM is an intuitive criterion that captures the case of there being a fixed underlying network. Second, as we now discuss, we can test LLM using a natural extension of our consistently differentiates condition. Recall our story about a set of vectors $\{b^N\}_{N \in N_i}$ behaviorally differentiating agent $i$. In the case of GLM, we asked that these vectors aligned with agent $i$ in each group, differentiated agent $i$ from every other agent in a given group, and were consistent across groups. When moving from GLM to LLM, we are moving to a model where the social influence parameters $\pi_i^N$ and $\pi_i^M$ are actually connected across groups. Our condition for testing LLM extends our consistently differentiating condition taking this across-group connection into account. Specifically, we weaken differentiation so that, for each agent $j\neq i$, agent $i$ needs only to be differentiated from agent $j$ on average across all groups that they share.

\begin{definition}
     We say that a set of vectors $\{b^N\}_{N \in \mathcal{N}_i}$ with $b^N \in \mathbb{R}^X$ for each $N \in \mathcal{N}_i$ \textbf{weakly behaviorally differentiates} agent i if the following are satisfied.
     \begin{enumerate}
         \item (Alignment) For each $N \in \mathcal{N}_i$, $b^N \cdot p_i^N > 0$.
         \item (Weak Differentiation) For each $j \neq i$, $\sum_{N \in \mathcal{N}_i \cap \mathcal{N}_j} b^N\cdot (p_i^N - p_j^N) \geq 0$.
     \end{enumerate}
\end{definition}

\begin{definition}
    We say that a dataset $\{p^N\}_{N \in \mathcal{N}}$ \textbf{consistently weakly differentiates} agent $i$ if every $\{b^N\}_{N\in \mathcal{N}_i}$ which weakly behaviorally differentiates agent $i$ does not satisfy $\sum_{N \in \mathcal{N}_i}b^N \ll 0$.
\end{definition}

\begin{theorem}\label{Thm:LLMCharacterization}
    For a dataset $\{p^N\}_{N \in \mathcal{N}}$, the following are equivalent.
\begin{enumerate}
    \item $\{p^N\}_{N \in \mathcal{N}}$ is consistent with the Luce linear-in-means model.
    \item For all agents $i$, $\{p^N\}_{N \in \mathcal{N}}$ consistently weakly differentiates agent $i$.
\end{enumerate}
\end{theorem}

Observe that if our dataset consistently weakly differentiates each agent, then it consistently differentiates each agent. It then follows from Theorem \ref{Thm:GenChar} that the collection of sets $\{co^{-1}(\Delta(p^N),p_i^N)\}_{N \in \mathcal{N}_i}$ have a point of mutual intersection. In the case of GLM, this is the set of feasible $v_i$. The strengthening of differentiation to weak differentiation is exactly what guarantees us that the social influence parameters $\pi_i^N$ follow a Luce rule across groups.

\subsection{Uniform Model}\label{Sec:uniLuce}
We now turn our attention to the uniform model. Recall that, in this case, agent $j$ influences agent $i$ only through agent $j$'s contribution to the unweighted average outcome of all agents other than $i$ in the group. As mentioned earlier, this version of our model most closely corresponds to the original version of the linear-in-means model proposed by \citet{manski1993identification} in the sense that the uniform model allows for peer effects but does not allow for more complicated network effects. In this section, we show how the results and techniques from Section \ref{Sec:GLM} simplify and allow for easy testing and identification.

To begin, recall that, in the uniform model, $\pi_i^N(j)=(1-\alpha_i^N)\frac{1}{|N|-1}$. If we use $\bar{p}_{-i}^N$ to denote the unweighted average of $\{p_j^N\}_{j\neq i}$, we can rewrite Equation \ref{EQ:linInMeans} for the uniform model as 
\begin{equation}\label{EQ:uniLinInMeans}
    p_i^N=\alpha_i^N v_i +(1-\alpha_i^N) \bar{p}_{-i}^N.
\end{equation}
Equation \ref{EQ:uniLinInMeans} exactly mirrors Equation \ref{EQ:linInMeans} in the case of two agents. Further, the across group restrictions for ULM mirror those for the case of GLM. In particular, GLM puts no restrictions on how $\pi_i^N$ and $\pi_i^M$ are related, while ULM puts no restrictions on how $\alpha_i^N$ and $\alpha_i^M$ are related. With this observation in mind, testing and identification procedures for ULM effectively mirror those for GLM, restricted to the case of observing groups of two agents. To formalize this, we consider the set of feasible bliss points of agent $i$ in group $N$ for ULM.

\begin{equation}\label{EQ:UniInverseCone}
    co^{-1}(\bar{p}_{-i}^N,p_i^N)=\{v\in \Delta(X)|v = \gamma_i p_i^N + \gamma_{-i}\bar{p}_{-i}^N, \gamma_{-i} \leq 0, \gamma_i +\gamma_{-i} = 1\}
\end{equation}

Once again, Equation \ref{EQ:UniInverseCone} mirrors Equation \ref{EQ:InverseCone} in the case that Equation \ref{EQ:InverseCone} is considering a group of two agents. We can also naturally extend our consistently differentiating condition for testing GLM to the case of ULM.

\begin{definition}
     We say that a set of vectors $\{b^N\}_{N \in \mathcal{N}_i}$ with $b^N \in \mathbb{R}^X$ for each $N \in \mathcal{N}_i$ \textbf{behaviorally differentiates agent $i$ from the average} if the following are satisfied.
     \begin{enumerate}
         \item (Alignment) For each $N \in \mathcal{N}_i$, $b^N \cdot p_i^N > 0$.
         \item (Differentiation from the Average) For each $N \in \mathcal{N}_i$, $b^N\cdot (p_i^N - \bar{p}_{-i}^N) \geq 0$.
     \end{enumerate}
\end{definition}

\begin{definition}
    We say that a dataset $\{p^N\}_{N \in \mathcal{N}}$ \textbf{consistently differentiates agent $i$ from the average} if every $\{b^N\}_{N\in \mathcal{N}_i}$ which behaviorally differentiates agent $i$ from the average does not satisfy $\sum_{N \in \mathcal{N}_i}b^N \ll 0$.
\end{definition}

The only difference between our earlier behaviorally differentiating definition and the definition of behaviorally differentiating from the average is that, in this case, we ask for differentiation to hold against the average choice of other agents in group $N$ rather than against the choices of each agent in group $N$. We now state our characterization of the uniform linear-in-means model.

\begin{theorem}\label{Thm:UniChar}
For a dataset $\{p^N\}_{N \in \mathcal{N}}$, the following are equivalent.
\begin{enumerate}
    \item $\{p^N\}_{N \in \mathcal{N}}$ is consistent with the uniform linear-in-means model.
    \item For every $i \in \mathcal{A}$, the collection of sets $\{co^{-1}(\bar{p}_{-i}^N,p_i^N)\}_{N \in \mathcal{N}_i}$ has a point of mutual intersection.
    \item For all agents $i \in \mathcal{A}$, $\{p^N\}_{N \in \mathcal{N}}$ consistently differentiates agent $i$ from the average.
\end{enumerate}
\end{theorem}

\subsubsection*{Identification Properties}

Now that we can test for ULM, the natural question is can we identify the parameters of the uniform model. As we mentioned earlier, the identification properties of ULM mirror the identification properties of GLM when each group consists of two agents. In this sense, we can directly appeal to Corollary \ref{cor:PointIdent}, now asking for linear independence of $(p_i^N -p_{-i}^N)$ and $(p_i^M-p_{-i}^M)$, in order to recover agent $i$'s ideal point. Thus, in the uniform model, an analyst can generically recover an agent's ideal point by observing two groups containing that agent. Now turning to the recovery of the weight of peer effects, the identification properties of $\alpha_i^N$ in the uniform model are particularly nice. We can once again appeal to a result about GLM, in this case Proposition \ref{Prop:PiPointIdent}. However, since we are only dealing with $v_i$ and $p_{-i}^N$, our prior affine independence condition reduces to simply asking that $v_i$ and $p_{-i}^N$ are not equal. If this is the case, then we are able to recover $\alpha_i^N$ and, by extension, $(1-\alpha_i)$ which corresponds to the importance of agent $i$'s peers in their decision making process.

\section{Extensions}\label{Sec:Extensions}
In this section, we discuss how our results from Section \ref{Sec:Results} extend under alternative assumptions on our model and the data. Our first extension is to consider relaxing the assumption that our data satisfies $\sum_{x \in X} p_i^N(x)=1$ (as well as $p_i^N \geq 0$). In this case, all of our results extend (mostly) unchanged. In our second extension, we relax the assumption that agents are only influenced by way of conformism. That is to say, we allow for $\pi_i^N(j)$ to be negative when $j \neq i$ as well as $\sum_{j \in N}\pi_i^N(j)$ to not be equal to one. In this case our identification and geometric testing results go through mostly unchanged. For our third extension, we consider three additional forms of variation and show how they can provide even stronger results than what we have discussed in Section \ref{Sec:Results}. Finally, our last extension considers how to recover identification of network effects in the case when there are more agents than there are outcomes.

\subsection{Choices in Euclidean Space}

Thus far, we have taken the perspective of stochastic choice.  Agents in our model choose a probability distribution.  However, nothing precludes us from investigating more sophisticated choice environments. The classic one-dimensional framework has as its choice space $[0,1]$.  The unit interval can be interpreted in many ways---any cardinal unidimensional choice can be modeled this way.  We have interpreted the unit interval as a probability distribution over two objects, leading to the simplex as the natural generalization.  Similarly, if we were to imagine an individual deciding how to spend their day, $[0,1]$ might measure the proportion of hours devoted to sleep vs. waking, whereas a model with three alternatives might further refine waking hours into work and leisure.

However, suppose that $[0,1]$, or equivalently $[0,100]$, represents the score on a test. We might also be interested in scores on multiple exams:  a member of $[0,1]^2$ might reflect the scores on an English language exam and on a Turkish language exam. In fact, Equation~\eqref{EQ:linInMeans} is meaningful whenever there is a convex subset of a vector space $Y$, and each $p_i^N$ lies in $Y$.  The particular case of $Y=\Delta(X)$ is only one example.   All three of our particular models (general, Luce, and uniform) also have ready interpretation in terms of more general vector spaces.

Let us illustrate this point via example, whereby our choice space is $Y=[0,100]^T$.  We can think of $T$ as indexing a collection of tests.  Any $p_i^N\in Y$, where $p_i^N$ no longer necessarily represents probabilities is a vector of test scores.  Our equation $p_i^N = \pi_i^N(i)v_i+\sum_{J\in j\in N\setminus i}\pi_i^N(j)p_j^N$ asserts that an individual's test scores are an average of some ``baseline scores'' ($v_i$) and the scores of other individuals present.

In this more general setup, the definition of inverse cone remains unchanged and still captures the set of feasible $v_i$ given the data.  \begin{equation}\label{EQ:InverseConeGeneral}
    co^{-1}(Y,p_i^N)=\{v\in Y|v = \sum_{j \in N}\gamma_j p_j^N, \gamma_j \leq 0 \text{ } \forall j \in N\setminus i, \sum_{j \in N}\gamma_j = 1\}
\end{equation}
Theorem \ref{Thm:GenChar} parts 1 and 2 remain equivalent.  More generally, the duality conditions (such as our consistently differentiating condition) will change from domain to domain, but can be characterized for polyhedral domains.  The downside is that, outside of the simplex domain, these dual conditions become more difficult to interpret. In regards to identification, since $co^{-1}(Y,p_i^N)$ still captures the set of feasible $v_i$, all of the results from Section \ref{Sec:Ident} go through unchanged.

% One remaining caveat is that, in more general vector spaces, we may allow for the possibility that individual weights need not sum to one. As mentioned in Section \ref{Sec:Model}, this causes problems for many of our results. For example, if $Y= \mathbb{R}^M_+$, then it is mathematically meaningful to allow any weights $\pi_i^N$ for which $\pi_i^N(j)\geq 0$ for all $j$ and $\pi_i^N(i)>0$.  Which sets of weights might be mathematically meaningful depends, in general, on the domain under consideration. The economic interpretation will likely dictate which set of weights to choose.

\subsection{Non-Conformism Social Influence}\label{sec:negweights}
Up until now, we have focused on the case of social influence via conformism. That is, we have focused on the case when $\pi_i^N$ corresponds to convex weights. However, there are cases where one person might influence another to act differently than themselves. As an example, consider an unruly teenager who observes the actions of their parents. In an act of defiance or in order to differentiate themselves from their parents or some of their peers, the teenager aims to choose an action that differs from their parents or peers. We can model this by allowing $\pi_i^N(j)$ to be negative. In this section, we consider the extension of our model and main results to the case of (potentially) negative weights.

In this section, suppose that $Z \subseteq \mathbb{R}^X$ corresponds to the set of feasible bliss points, $v_i$, for each agent as well as the set of potential outcomes, $p_i^N$, for each agent.\footnote{As one example, $Z$ could correspond to $\Delta(X)$, in which case we return to the data type we considered in Sections \ref{Sec:Model} and \ref{Sec:Results}.} For our analysis of the negative weight case, we make a single assumption on $\pi_i^N$ which is $$\pi_i^N(i)\in (0,1].$$ We treat this assumption as conceptually fundamental as it says that an agent cares about their own ideal point. Other than that, we make no restrictions on $\pi_i^N$ either within or across groups. This also means that we are dropping the assumption of $\sum_{j \in N}\pi_i^N(j)=1$. We still use Equation \ref{EQ:linInMeans} to govern each agent's choice. Despite the weakening of so many of our assumptions, we are still able to develop the set which corresponds to an agent's set of feasible bliss points. \begin{equation}\label{eq:FeasibleVNeg}
    co^{-1}(Z,p_i^N)=\{v\in Z|v=\sum_{j\in N}\gamma_jp_j^N, \gamma_i\geq 1\}
\end{equation}
Just as was the case in the previous section, conditions 1 and 2 from Theorem \ref{Thm:GenChar} go through using $co^{-1}(Z,p_i^N)$ instead of $co^{-1}(\Delta(p^N),p_i^N)$. However, unlike the previous section, a version of the duality results need not go through. It turns out that, for general $Z$, $co^{-1}(Z,p_i^N)$ is not a polyhedral set, and thus is not amenable to the duality techniques we use to prove condition 3 in Theorem \ref{Thm:GenChar}.\footnote{Specifically, when $Z=\mathbb{R}^X$, it turns out that $co^{-1}(Z,p_i^N)$ corresponds to an open half-space, relative to the linear subspace formed by $\{p_i^N\}_{i \in N}$, plus the point $p_i^N$.}

As for identification, we have to make one minor change to Theorem \ref{thm:pointIdent} in order for it to go through. Once we make this change, Theorem \ref{thm:pointIdent} and its proof go through unchanged.
\begin{theorem}\label{thm:NegPointIdent}
    Fix agent $i$ and suppose that $\{p^N\}_{N \in \mathcal{N}}$ is consistent with GLM under the assumptions of this section (i.e. with arbitrary $\pi_i^N(j)$). If, for each non-zero vector $p$ in the span of $\{p_j^N\}_{N \in \mathcal{N}_i,j \in N}$, there is some group $M \in \mathcal{N}_i$ such that $p$ is not in the span of $\{p_j^M\}_{j \in M}$, then $v_i$ is point identified.
\end{theorem}
Theorem \ref{thm:NegPointIdent} relaxes Theorem \ref{thm:pointIdent} in two distinct ways. First, we relax the assumption that $\pi_i^N(j)\geq 0$, allowing for potentially negative influence. Second, we allow for the sum of weights, $\sum_{j \in N} \pi_i^N(j)$, to not sum to one. Going from Theorem \ref{thm:pointIdent} to Theorem \ref{thm:NegPointIdent}, we go from considering vectors of the form $p_i^N-p_j^N$ to vectors of the form $p_j^N$. This change is due to the fact that we allow for any linear combination of $\{p_j^N\}_{j \in N\setminus i}$ to impact agent $i$'s choice.

\subsection{Other Forms of Variation}

Throughout this paper, we have focused on group variation being the main source of variation in the data. We now consider three alternative forms of variation; network variation, characteristic variation, and product-attribute variation. We discuss each of these forms of variation in a series of three remarks.

\begin{remark}[Network Variation]
    Instead of (or in addition to) group variation, we could consider variation of the underlying network. That is to say, fix a group $N$ and consider two realizations of some observable variable, such as time period, or regime, $R$ and $R'$. Suppose that an agent's social influence parameters are given by $\pi_i$ when $R$ is observed and $\pi'_i$ when $R'$ is observed. Neither $\pi_i$ nor $\pi'_i$ are observable to the analyst. If no restrictions are placed on $\pi_i$ or $\pi'_i$, testing for the linear-in-means model with network variation can be done using our Theorem \ref{Thm:GenChar}. Further, all of the identification results of Section \ref{Sec:Ident} also extend to this arbitrary network variation setup.
\end{remark}

\begin{remark}[Individual-Characteristic Variation and the Reflection Problem]

In our baseline setup, we assume that no individual-level characteristics are observed beyond agent identity. A central component of the reflection problem, as emphasized by \citet{manski1993identification}, is the challenge of disentangling exogenous effects and correlated effects from endogenous group effects. We can use a two step procedure that allows us to differentiate between agents who choose similarly due to peer influence (endogenous effects) and those agents who choose similarly due to the agents themselves being similar (correlated effects).

In the first step, we apply the identification strategy developed in Section \ref{Sec:Ident}, which enables recovery of the latent variables $v_i$ for each agent. In the second step, we treat $v_i$ as the dependent variable and conduct a standard analysis relating $v_i$ to observed characteristics. Under this framework, the first stage captures endogenous effects associated with the reflection problem, while the second stage isolates how characteristics affect each agent's ideal point.

Before moving on to our next remark, we take a moment to discuss the connection between our model, our results, and the reflection problem of \citet{manski1993identification}. The closest version of our model to the model studied by the reflection problem is our uniform model with individual-characteristic variation. Both models ask that an agent is influenced by the unweighted average of the other agents in their group. This corresponds to the endogenous effects of the reflection problem. Notably, we are able to identify endogenous effects in our setup.

The reflection problem has two further important effects, correlated effects and exogenous effects. In the setting of \citet{manski1993identification}, an agent exhibits correlated effects if their outcome depends on which group(s) that agent belongs. Our model has no such term. This, however, is partly due to the difference in how \citet{manski1993identification} and our model treats group belonging. In the setup of Manski, group belonging is a fixed characteristic of an agent, while, in our setup, an agent's group necessarily varies. Since our agents' ideal points are assumed to be invariant of their group, the exact type of correlated effect described in Manski is not present in our model.\footnote{In theory, one could encode the list of groups to which an agent belongs in our setup as an independent variable and then analyze the relation between $v_i$ and this variable. This would be an alternative way of encoding correlated effects as considered in \citet{manski1993identification}.} However, we believe the essence of a correlated effect is the situation in which two agents with similar characteristics (potentially observed or unobserved) choose similar outcomes due to them having similar ideal points. In this sense, our model allows for a notion of correlated effects which differs from the original model of Manski. Given our identification results, we can recover $v_i$ and $v_j$ for two agents $i$ and $j$ as well as analyze the relationship between characteristics, $v_i$, and $v_j$. While we don't propose a specific measure to quantify these relationships, the relationship between $v_i$, $v_j$, as well as the characteristics of $i$ and $j$ exactly correspond to what we believe is the essence of a correlated effect.

The final type of important effect in the reflection problem correspond to exogenous effects. In Manski's setup, an agent's outcome is allowed to depend on the average \textit{characteristics} of other agents in their group. Once again, since our base-model has no notion of characteristics, our model has no term capturing exogenous effects. Further, there is no way to straightforwardly extend our results to capture the notion of exogenous effects considered in \citet{manski1993identification}. In particular, the presence of meaningful exogenous effects would mean that the ideal points of our agents would vary across groups, depending on the characteristics of the other agents in these groups. Our results crucially rely on the fact that an agent's ideal point does not vary with their group.\footnote{Just as was the case with correlated effects, we could treat the average characteristics of agents in \textit{every} group to which agent $i$ belongs as an independent variable that impacts $v_i$. This would capture a notion of exogenous effect that still allows for our techniques.}

\end{remark}

\begin{remark}[Product-Attribute Variation]
    Another variation that could be introduced in our framework is product attribute variations. Suppose that there is a set of $K$ observable attributes, with $a_x$ denoting the vector of attributes for product $x$. $a_x$ includes not only things that affect product quality, but also things like price, advertising, etc. We assume that the ideal point of each agent is a linear function of observable product attributes. This is captured by an agent-specific real vector $\beta_i$ such that $u_i(x) = \beta_i a_x$ for each $i$ and $x$. This means that we assume that attributes affect utility in the same way for all products. Then the ideal point is calculated as $v_i(x)=\frac{u_i(x)}{\sum_{y\in X} u_i(y)}$. Identification of $\beta_i$ then proceeds with a two-step procedure. As above, we first recover $v_i$ for each agent. In the second step, we treat $v_i$ as the outcome variable and proceed with standard analysis of the relationship between $v_i$ and observed attributes.
\end{remark}

\subsection{Large Groups and Small Choice Sets}

In Section \ref{Sec:Ident}, our ability to recover network effects relied on there being more alternatives in $X$ than there are agents in a group, $|X| \geq |N|$. This imposes that we are considering an environment where there are many outcomes but few agents. In practice, this may not always be the case. However, our results still suggest a method for handling the case when there are fewer alternatives than there are agents, $|X| < |N|$. This method requires the analyst to either make assumptions or have partial information about the underlying network structure before doing their analysis.

We now consider an assumption on the underlying network structure that allows for easy adaptation of our techniques. Fix an agent $i$ and suppose that agent $i$ can sort all other agents into types, $t \in T_i$. Note that the set of types is allowed to depend on the agent $i$. Further, the assignment of agent $j$ to a type can also depend on the reference agent $i$ as well as the group $N$ under consideration. We make two assumptions on this set of types. First, we ask that, if agents $j$ and $k$ are both type $t$, then $\pi_i^N(j)=\pi_i^N(k)$. By asking that two agents of the same type have the same $\pi_i$, we are able to do the same reduction as we did with the uniform model of Section \ref{Sec:uniLuce}, allowing us to consider the unweighted average choice of agents in group $t$ rather than considering the choice of each agent in a group separately. Second, we ask that $|T_i| \leq |X|-1$. By imposing this, each group $N$ will have at most $|X|$ relevant data points driving network effects, agent $i$'s choice and the average choice of each type $t$. These two assumptions together allow us to implement all of our results from Section \ref{Sec:GLM} using the average choices of each type as our data rather than the choices of each agent.

Without any prior knowledge about agents, one could argue that the process of assigning agents to types may not be well-founded. However, we argue that analysts often have access to data that can assist in this assignment procedure. In particular, suppose an analyst is studying college outcomes and has access to data about students. The dataset records each student's roommate, dormitory, classes, and major. With this information, the analyst could sort agents into types in the following manner. Fixing a student, which we will denote as agent $i$, types correspond to a binary vector in $\mathbb{R}^4$ where each dimension records if the student is (1) agent $i$'s roommate, (2) in the same dormitory as agent $i$, (3) in the same class(es) as agent $i$, and (4)  in the same major as agent $i$. Thus, if the analyst has partial information about the agents or the underlying network, they can develop a natural set of types that leads to the assumptions made in the previous paragraph.

\section{Related Literature}\label{Sec:Discussion} 
In this paper we study the testing and identification properties of the linear-in-means model when an analyst observes group variation. As part of our analysis, we show that the identification of endogenous effects, as is considered in the reflection problem of \citet{manski1993identification}, is a generic problem when each agent's decision space is one dimension but stops being generic when we move to higher dimensions. This highlights the identifying power of more granular data in the linear-in-means model and we emphasize this as a key takeaway of our analysis. An obvious followup to our work would be the empirical implementation of our methods. While we have provided theoretical identification results, there is further work needed in order to bring our results to real, finite data. We leave this to future research. We now conclude with a discussion of this paper's place in the related literature.

Our paper is related to several strands of literature. We begin by discussing the strand which studies the linear-in-means model of social interactions. Predating the modern literature on the linear-in-means model, \citet{keynes1937general} considers a model of financial markets via a story of beauty contests. In this setting, an agent wishes to take the action that coincides with the average action of the rest of the population. In our setup, this corresponds to $\pi_i^N(i)=0$ and $\pi_i^N(j)=\frac{1}{|N|-1}$. More recently, \citet{ushchev2020social} studies the microfoundations and comparative statics of the linear-in-means model allowing for arbitrary network structure. As mentioned earlier, \citet{ushchev2020social} along with \citet{blume2015linear}, \citet{boucher2016some}, and \citet{kline2020econometric} show that the linear-in-means choice rule can be achieved as the best response to a quadratic loss utility function in a complete information game where each agent knows each $v_i$ and $\pi_i^N$. \citet{golub2020expectations} consider an extension of this setup where agents have incomplete information and relate the linear-in-means model to higher-order expectations as well as conventions in networks. We build on this literature by studying the empirical content of the linear-in-means model.

More closely related to our paper is the strand of literature that focuses on studying the identification properties of the linear-in-means model as well as other related models of peer effects. A seminal contribution in this literature is \citet{manski1993identification} and his discussion of the reflection problem. Important to our analysis is the following takeaway from the reflection problem of \citet{manski1993identification}. It is in general difficult to identify the social influence of a group on an agent due to the endogenous nature of outcomes. In our setting, this corresponds to identifying both the underlying network structure and the corresponding weights on directed edges in this network. Much of the literature following \citet{manski1993identification} aims to identify social interaction parameters when the underlying network structure is (partially) known. This literature is extensive, so we list the following, all of which provide various conditions in order to recover identification of social interaction parameters in linear-in-means style models; \citet{lee2007identification}, \citet{graham2008identifying}, \citet{bramoulle2009identification}, \citet{de2010identification}, \citet{blume2011identification}, \citet{boucher2014peers}, \citet{blume2015linear}, \citet{de2017econometrics}. More recently, \citet{boucher2024toward} extends the linear-in-means to a CES in means model of social influence and provides identifications in their setting.

Perhaps most closely related to our analysis within this literature is the work of \citet{lewbel2023social} and \citet{de2024identifying}. Both of these papers study the joint identification of social interaction parameters as well as the identification of the underlying network structure.\footnote{\citet{breza2020using} and \citet{Griffith2023} also consider joint identification of peer effects and network structure. However, \citet{breza2020using} asks for a stronger form of data where the analyst knows the distribution over how many connection each agent has to other agents with specific characteristics. Similarly, \citet{Griffith2023} makes assumptions on the underlying network density to recover the unobserved network structure. \citet{battaglini2022endogenous} also estimates network connections in a related model of endogenous network formation.} While \citet{lewbel2023social} utilizes group variation and \citet{de2024identifying} utilizes long-run panel data, a common tool for identification in each of these studies is the use of characteristic variation in order to recover identification. The models of these two papers also differ in the fact that the characteristics of each agent show up in the choice procedure of other agents. Adapting to our context, the core components of these models can be summarized by\begin{equation}\label{eq:LIMwithChar}
    p_i^N=\pi_i^N(i)v_i + \alpha z_i + \beta\sum_{j \in N \setminus i} \pi_i^N(j)p_j^N + \gamma \sum_{j \in N \setminus i} \pi_i^N(j) z_j+\delta_N
\end{equation} where $z_i$ is the characteristic vector of agent $i$, $\alpha$ is the average impact of characteristics on each agent's bliss point, $\beta$ corresponds to the average impact of other agents' choice on each agent's choice, $\gamma$ corresponds to the average impact of other agents' characteristics on each agent's choice, and $\delta_N$ is a group or network specific effect. \citet{de2024identifying} interprets $\beta$ as the endogenous peer effects, $\gamma$ as the exogenous peer effects, and $\delta$ as the correlated effects from the reflection problem of \citet{manski1993identification}. Equation \ref{eq:LIMwithChar} also reflects the nature of the main specification of \citet{manski1993identification} once we assume that $\pi_i^N(j)=\pi_i^N(k)$ for all $j,k\neq i$. Our analysis differs from this literature in two ways. First, our analysis focuses on the study of simultaneity problem of endogenous effects. Second, we require no knowledge of characteristics (other than identity) to jointly recover each agent's bliss point as well as the underlying network structure. The key to our identification result is that our outcome vector $p_i^N$ is multidimensional. In fact, generically, our results show that identification fails in our setup when the outcome variable is one-dimensional.

Our paper also contributes to the choice theoretic literature studying social interactions. To our knowledge, this literature begins with \citet{cuhadaroglu2017choosing}, who studies a two-period model with social influence taking effect in the second period. \citet{borah2018choice} also considers a two-period model of social influence. In the first period, social influence is used to form an agent's consideration set, and the second period is used for choice. \citet{kashaev2023peer} consider a model of social influence where the choices of an agent's peers directly form their consideration set. One of the main goals of this work is the actual identification of underlying social parameters, which they are able to achieve through dynamic choice variation. \citet{bhushan2023beliefs} considers a model of choice where agents influence each other through their beliefs. An agent's beliefs are determined through a process similar to ULM. Choices then correspond to subjective expected utility given these beliefs. Most closely related to our work in this literature is the work of \citet{chambers2023behavioral} who consider a version of the linear-in-means model. They focus on a setting with menu variation (i.e., variation of $X$) with a fixed group and network structure. This differs from our analysis in that our focus is on group variation, and we can accommodate network variation.

More generally, our paper is related to the literature studying the empirical content of strategic settings. Part of this literature focuses on testing the empirical content of specific solution concepts. \citet{sprumont2000testable} studies the testable content of Nash equilibrium. The work of \citet{haile2008empirical} finds that the quantal response equilibrium has no empirical content. Similarly, \citet{bossert2013every} and \cite{rehbeck2014every} find that backwards induction has no empirical content when we only observe the induced choice function. Another portion of this literature focuses on characterizing the empirical content of specific (types of) games. To begin, \citet{lee2012testable} characterizes the testable content of zero-sum games. \citet{carvajal2013revealed} provides a revealed preference style characterization of the Cournot model of competition. Finally, \citet{lazzati2023ordinal} characterizes the empirical content of Nash equilibrium in monotone games. As mentioned earlier, the linear-in-means model can be thought of as arising from Nash equilibrium play where each agent's utility is given by Equation \ref{EQ:baseutility}. As a result of this equivalence, we characterize Nash equilibrium play in the corresponding game when we observe player variation.

Finally, our paper is also related to the literature on stochastic choice studying agents who have preferences for non-deterministic bundles. The idea of deliberately stochastic preferences goes back to at least \cite{machina1985stochastic}. Recently, \citet{cerreia2019deliberately} axiomatizes data that arises from agents choosing with deterministic preferences over lotteries. Similarly, \citet{fudenberg2015stochastic} characterizes stochastic choice data that arises from agents who have cardinal preferences over each alternative but face a perturbation to their utility function within the simplex. \citet{allen2019identification} studies the identification properties of a similar class of perturbed utility functions. There has been little work on these types of perturbed utility functions with social influence components. \citet{hashidate2023social} is an exception and considers a perturbed utility function where the perturbation corresponds to a norm. Recall that the choices in the linear-in-means model can be induced by a perturbed utility model where an agent's base utility is given by a quadratic loss function with reference to their ideal point. The perturbation then corresponds to a sum of quadratic loss functions with each one referencing another agent's choice.

\newpage 
\bibliographystyle{ecta}
\bibliography{linfluence}

\newpage

\appendix

{\begin{center} {\huge{\textsc{Appendix}}}\end{center} }

\section{Preliminary Results}\label{app:Prelim Resulits}

We begin with a preliminary lemma that formalizes our discussion at the start of Section \ref{Sec:GLM}.

\begin{lemma}\label{inverseconelemma}
    $p_i^N = \gamma_i v_i + \sum_{j \in N \setminus i} \gamma_j p_J^N$ for $v_i \in \Delta(X)$ with $\gamma_j \geq 0$, $\gamma_i > 0$, and $\sum_{j \in N}\gamma_j=1$ if and only if $v_i \in co^{-1}(\Delta(p^N),p_i^N)$.
\end{lemma}

\begin{proof}
    Suppose that $p_i = \gamma_i v_i + \sum_{j \in N \setminus i} \gamma_j p_J^N$ for $v_i \in \Delta(X)$ with $\gamma_j \geq 0$ and $\sum_{j \in N}\gamma_j=1$. Since $0 < \gamma_i \leq 1$, we can recover that $v_i = \frac{1}{\gamma_i} p_i^N + \sum_{j \in N \setminus i}\frac{-\gamma_j}{\gamma_i}p_j^N$. Since $\gamma_j \geq 0$ and $\sum_{j \in N}\gamma_j=1$, $\frac{-\gamma_j}{\gamma_i} \leq 0$ and $\frac{1}{\gamma_i} + \sum_{j \in N \setminus i}\frac{-\gamma_j}{\gamma_i}=1$ and so $v_i \in co^{-1}(\Delta(p^N),p_i^N)$.

    Now suppose that $v_i \in co^{-1}(\Delta(p^N),p_i^N)$. This tells us that $v_i = \sum_{j \in N}\gamma_j p_j^N$ with $\gamma_j \leq 0$ $\forall j \in N\setminus i$ and $\sum_{j \in N}\gamma_j = 1$. It follows that $\gamma_i \geq 1$. By basic algebra, we can solve for $p_i$ and get $p_i=\frac{1}{\gamma_i} v_i + \sum_{j \in N \setminus i}\frac{-\gamma_j}{\gamma_i}p_j^N$. Since $\gamma_j \geq 0$ for $j \in N\setminus i$ and $\sum_{j \in N}\gamma_j=1$, $\frac{-\gamma_j}{\gamma_i} \geq 0$ and $\frac{1}{\gamma_i} + \sum_{j \in N \setminus i}\frac{-\gamma_j}{\gamma_i}=1$ and so $p_i = \pi_i v_i + \sum_{j \in N \setminus i} \pi_j p_J^N$ for $v_i \in \Delta(X)$ with $\pi_j \geq 0$, $\pi_i > 0$, and $\sum_{j \in N}\pi_j=1$.
\end{proof}

Define $X_1 = \{x\in\mathbb{R}^n:\sum_i x_i = 1\}$.  So $X_1$ is not the simplex but rather the hyperplane containing the simplex. The following very simple result is closely related to classical results, for example \citet{girsanov}, Lemma 3.11, or characterizations of Pareto optimality, relating to Negishi weights \emph{e.g.} \citet{smale1975,wan1975local} but is written to be close in form to \citet{billot,samet1998common}.  The main difference from the previous pair of papers is that no compactness assumptions are claimed, but rather polyhedrality of the sets is required.

\begin{theorem}\label{thm:polyhedral}Let $A_1,\ldots,A_k$ each be nonempty subsets of $X_1$, and suppose additionally that each $A_i$ is a polyhedron.  Suppose that $\bigcap_{i=1}^k A_i = \varnothing$.  Then the following are equivalent:
\begin{enumerate}
\item $\bigcap_{i=1}^k A_i = \varnothing$.
\item For each $i=1,\ldots,k$, there are $p_i \in \mathbb{R}^n$ for which $\sum_{i=1}^k p_i = 0$ and such that $p_i \cdot x_i > 0$ for all $x_i\in A_i$.
\end{enumerate}
\end{theorem}

\begin{proof}We show that 1 implies 2.  Observe that $X_1$ is itself defined by a finite set of linear inequalities. We now consider $Y=\{(x,\ldots,x):x\in X_1\}\subseteq \mathbb{R}^{nk}$ and $\prod_{i=1}^k A_i$.  These sets are both clearly polyhedra. Further, by hypothesis, $Y\cap \prod_{i=1}^k A_i=\varnothing$.  By Corollary 19.3.3 and Theorem 11.1 of \citet{rockafellar}, there exists $(p_1,\ldots,p_k)\in\mathbb{R}^{nk}$ and $c\in\mathbb{R}$ for which for all $x\in X_1$, $\sum_i p_i \cdot x < c$ and for all $(x_1,\ldots,x_k)\in\prod_i A_i$, $\sum_i p_i\cdot x_i > c$.  In particular we may choose $c$ so that $\inf_{(x_1,\ldots,x_k)\in \prod_i A_i}\sum_i p_i\cdot x_i > c$.

Define $p_i^* = p_i - (c/k)\mathbf{1}$ and observe that, because all $A_i$ are subsets of $X_1$, the $c$ in the inequalities gets replaced with $0$ when replacing $p_i$ with $p_i^*$. Specifically, this also tells us that (by considering indicator functions of the form $\mathbf{1}_j$ for $j\in\{1,\ldots,n\}$), we have $\sum_i p_i^* \ll 0$.  Further, it tells us that for each $x,y\in X$, $\sum_i p_i^*(x)(k+1)-\sum_i p_i^*(y)k < 0$, as $(k+1)\mathbf{1}_x-k\mathbf{1}_y\in X_1$.  Thus $\sum_i \frac{(k+1)p_i^*(x)-kp_i^*(y)}{2k+1}<0$, and by taking limits, $\sum_i p_i^*(x) \leq \sum_i p_i^*(y)$.  Since $x$ and $y$ are arbitrary, this implies that $\sum_i p_i^*(x)=\sum_i p_i^*(y)$ for all $x,y\in X$.

Now, define for each $i=1,\ldots k$, $c_i = \inf_{x_i \in A_i}p_i^* \cdot x_i$.  Then we have $\sum_i c_i > 0$. Fix $\epsilon > 0$ so that $\sum_i (c_i - \epsilon) > 0$.  For each $i$ then let $q_i^* = p_i^* - (c_i-\epsilon)\mathbf{1}$ and observe that for all $x_i \in A_i$, $q_i^*\cdot x_i = p_i^*\cdot x_i -(c_i -\epsilon)\cdot x_i\geq\epsilon>0$ as $x_i \in X_1$.  Further, $\sum_i q_i^* \ll 0$ remains valid as $\sum_i(c_i-\epsilon) > 0$.  Further, it remains true that $\sum_i q_i^*(x) = \sum_i q_i^*(y)$ for all $x,y\in X$ as we have simply subtracted a constant from each $q_i^*$.

We show that 2 implies 1.  Suppose by means of contradiction that there is $x^*\in\bigcap_{i=1}^k A_i$.  Then $p_i \cdot x^* > 0$ for all $i$ and in particular then $(\sum_{i=1}^kp_i)\cdot x^* > 0$.  But this is a contradiction as $(\sum_{i=1}^k p_i)\cdot x^* = 0\cdot x^* =0$.  \end{proof}

\section{Omitted Proofs}\label{app:Proofs}

\subsection{Proof of Theorem \ref{Thm:GenChar}}
Our proof of the equivalence between conditions (1) and (2) utilizes Lemma \ref{inverseconelemma} and follows quickly from it. Our proof of the equivalence between (1) and (3) utilizes Theorem \ref{thm:polyhedral}. For the separation, let us consider the sets $\Delta(X)$ and, for each $N$, $\hat{co}^{-1}(\Delta(p^N),p_i^N)$ where \begin{equation}\label{EQ:InverseConehat}
    \hat{co}^{-1}(\Delta(p^N),p_i^N)=\{v|v = \sum_{j \in N}\gamma_j p_j^N, \gamma_j \leq 0 \text{ } \forall j \in N\setminus i, \sum_{j \in N}\gamma_j = 1\}.
\end{equation} Note that this differs from Equation \ref{EQ:InverseCone} as $v$ is not required to be in the simplex. Owing to \citet{rockafellar} Theorem 19.3, each $\hat{co}^{-1}(\Delta(p^N),p_i^N)$ is polyhedral. 

\begin{proof}
We begin with the equivalence between (1) and (2). Suppose that $\{p_i\}$ is data consistent with GLM. Then, for each $i \in \mathcal{A}$, there exist $v_i$ and $\pi_i^N$ for each $N \in \mathcal{N}_i$ such that Equation \ref{EQ:linInMeans} holds for each $(i,N)$, Since $v_i$ is fixed across each $N \in \mathcal{N}$, by Lemma \ref{inverseconelemma}, the collection of sets $\{co^{-1}(\Delta(p^N),p_i^N)\}_{N \in \mathcal{N}_i}$ must have a point of common intersection.

Now suppose that, for each $i \in \mathcal{A}$, the collection of sets $\{co^{-1}(\Delta(p^N),p_i^N)\}_{N \in \mathcal{N}_i}$ has a point of common intersection. By Lemma \ref{inverseconelemma}, the $v$ in the intersection of $\{co^{-1}(\Delta(p^N),p_i^N)\}_{N \in \mathcal{N}_i}$ works as a feasible $v_i$ for each $N \in \mathcal{N}_i$. Specifically, for each $N \in \mathcal{N}_i$, there are $\gamma_j \geq 0$, with $\gamma_i >0$ and $\sum_{j \in N} \gamma_j =1$, such that $p_i^N= \gamma_i v + \sum_{j \in N \setminus i} \gamma_j p_j^N$. Take these $\gamma_j$ as $\pi_i^N(j)$. These $v$ and $\pi_i$ correspond to a GLM representation of the dataset.

Now we prove the equivalence between (1) and (3). By definition, data satisfying our consistently differentiating condition is equivalent to the lack of existence of, for each agent $i$, a $\{b^N\}_{N \in \mathcal{N}_i}$ which behaviorally differentiates $i$ and satisfies $\sum_{N \in \mathcal{N}_i} b^N \ll 0$. This is then equivalent to the following. For every $i \in \mathcal{A}$, there does not exist a collection of vectors $\{b^N\}_{N \in \mathcal{N}_i}$ with $b^N \in \mathbb{R}^X$ and $\sum_{N \in \mathcal{N}_i} b^N\ll 0$ such that for all $N\in N_i$, $b^N \cdot p_i^N >0$ and for each $j\neq i$ and $N$ for which $j\in N$, $b^N \cdot (p_i^N-p_j^N)\geq 0$. If there are $b^N$ as in the statement of the theorem, then the model is not valid.  Suppose, by means of contradiction that there are such $b^N$ but the model is valid.  Because $\pi_i^N(i)>0$, we know that $v_i = \frac{p_i^N}{\pi^N_i(i)}-\sum_{j\neq i}\frac{\pi_i^N(j)}{\pi_i^N(i)}p_j^N$.  In particular, since $\sum_x v_i(x) = 1$, we must have $\frac{1}{\pi^N_i(i)}-\sum_{j\neq i}\frac{\pi_i^N(j)}{\pi_i^N(i)}=1$.  Let $\lambda_i = \frac{1}{\pi^N_i(i)}$ and $\lambda_j= \frac{\pi_i^N(j)}{\pi_i^N(i)}$, then we have $\lambda_i-\sum_{j\neq i}\lambda_j = 1$, where each $\lambda_k \geq 0$.  
Then $\lambda_i p_i^N - \sum_{j\neq i}\lambda_j p_j^N = (\lambda_i-\sum_{j\neq i} \lambda_j)p_i^N+\sum_{j\neq i}\lambda_j(p_i^N-p_j^N)$.  Consequently $b^N\cdot v_i = (\lambda_i -\sum_{j\neq i}\lambda_j)b^N \cdot p_i^N +\sum_{j\neq i}\lambda_j b^N \cdot (p_i^N-p_j^N)>0$. So for each $N$, we have $b^N \cdot v_i>0$.  Consequently $(\sum_N b^N)\cdot v_i > 0$, but $\sum_N b^N \ll 0$,  a contradiction.  

On the other hand, suppose the model is violated.  Recall $\hat{co}^{-1}(\Delta(p^N),p_i^N)$. Observe that every $\gamma$ in this set satisfies $\gamma_i>0$.  Further, observe that $\hat{co}^{-1}(\Delta(p^N),p_i^N)$ reflects the set of possible $v_i$ which are compatible with the observed choice $p_i^N$.  That is, if there were some $v_i$ common to all $\hat{co}^{-1}(\Delta(p^N),p_i^N)$, we could define $\pi_i^N(i)=\frac{1}{\gamma_i}$ and $\pi_i^N(j)=-\frac{\gamma_j}{\gamma_i}$.  Now, by Theorem 19.3 of \citet{rockafellar}, each $\hat{co}^{-1}(\Delta(p^N),p_i^N)$ is polyhedral (owing to the fact that the set of $\gamma_j:j\in N$ satisfying the linear inequalities is a polyhedron). Since the model is violated, there is no $v_i$ common to all $\hat{co}^{-1}(\Delta(p^N),p_i^N)$. Consequently $\Delta(X) \cap \bigcap_{N:i\in N}\hat{co}^{-1}(\Delta(p^N),p_i^N) = \varnothing$.  By Theorem~\ref{thm:polyhedral}, there are weights $b^N$ and $q$ for which $q+\sum_N b^N = 0$, where $q \cdot x > 0$ for all $x\in\Delta(X)$, which means $q \gg 0$, and where $b^N \cdot p_i^N > 0$ (by taking a weight of one on $p_i^N$) and $b^N\cdot  ((k+1)p_i^N - kp_j^N) > 0$ for all $j$, which implies by taking limits with respect to $k$ and normalizing, $b^N\cdot (p_i^N-p_j^N) \geq 0$.  Finally, $\sum_N b^N = -q \ll 0$.  

\end{proof}

\subsection{Proof of Corollary \ref{cor:1DimChar}}

\begin{proof}
    By Theorem \ref{Thm:GenChar} condition (2), we know that a dataset is consistent with GLM if and only if, for each $i$, $\{co^{-1}(\Delta(p^N),p_i^N)\}_{N \in \mathcal{N}_i}$ have a point of common intersection. As mentioned in Section \ref{Sec:GLM}, if $p_i^N$ lies on the relative interior of $\Delta(p_{-i}^N)$, then $N$ offers no testable content for agent $i$. Specifically, if $p_i^N$ lies on the relative interior of $\Delta(p_{-i}^N)$ then $p_i^N$ can be written as a convex combination of $p_{-i}^N$ and $v \in \Delta(X)$ for any choice of $v$ with a vanishingly small weight put on $v$. It follows that we can restrict our attention to $co^{-1}(\Delta(p^N),p_i^N)$ for groups $N$ where $p_i^N \not \in \Delta(p_{-i}^N)$. These groups are exactly $\mathcal{N}_i^+\cup \mathcal{N}_i^-$. We have a series of observations for three cases.
    \begin{enumerate}
        \item Suppose $N \in \mathcal{N}_i^+\setminus \mathcal{N}_i^-$ \\
        In this case $p_i^N \geq p_j^N$ for all $j \in N\setminus i$ with a strict inequality for at least one $j$. Fix this $j$. In this case, set $\pi_i^N(k)=0$ for each $k \in N \setminus \{i,j\}$. It then follows that every $v$ satisfying $v \geq p_i^N$ can be rationalized by $p_i^N = \pi_i^N(i) v + \pi_i^N(j) p_j^N$ for some choice of convex weights, but any $v < p_i^N$ cannot.
        \item Suppose $N \in \mathcal{N}_i^-\setminus \mathcal{N}_i^+$ \\
        In this case $p_i^N \leq p_j^N$ for all $j \in N\setminus i$ with a strict inequality for at least one $j$. Fix this $j$. In this case, set $\pi_i^N(k)=0$ for each $k \in N \setminus \{i,j\}$. It then follows that every $v$ satisfying $v \leq p_i^N$ can be rationalized by $p_i^N = \pi_i^N(i) v + \pi_i^N(j) p_j^N$ for some choice of convex weights but any $v > p_i^N$ cannot.
        \item Suppose $N \in \mathcal{N}_i^+\cap \mathcal{N}_i^-$ \\
        In this case $p_i^N=p_j^N$ for each $j \in N$. In this case, any convex combination of $v$ with $p_j^N$ for $j \in N \setminus i$ which puts positive weight on $v$ would be different from $p_j^N$ if and only $p_i^N \neq p_j^N$. Thus, in this case we have that $v_i=p_j^N=p_i^N$.
    \end{enumerate}
    We now consider two cases.
    \begin{enumerate}
        \item Suppose that $\mathcal{N}_i^+\cap \mathcal{N}_i^- = \emptyset$. \\
        In this case $[0,\min_{N \in \mathcal{N}_i^-}p_i^N] = \bigcap_{N \in \mathcal{N}_i^-} \{co^{-1}(\Delta(p^N),p_i^N)\}_{N \in \mathcal{N}_i^-}$ and $[\max_{N \in \mathcal{N}_i^+,1}p_i^N,1] = \bigcap_{N \in \mathcal{N}_i^+} \{co^{-1}(\Delta(p^N),p_i^N)\}_{N \in \mathcal{N}_i^+}$. These two sets have a point of intersection if and only if $\max_{N \in \mathcal{N}_i^+}p_i^N \leq \min_{N \in \mathcal{N}_i^-}p_i^N$. The other two conditions are vacuous in this case.
        \item Suppose that $\mathcal{N}_i^+\cap \mathcal{N}_i^- \neq \emptyset$. \\
        By observation 3 above, for each $N^* \in \mathcal{N}_i^+\cap \mathcal{N}_i^-$, $co^{-1}(\Delta(p^{N^*}),p_i^{N^*})$ is exactly $p_i^{N^*}$ and so it must be the case that $v_i = p_i^{N^*}=p_i^=$ if the dataset is consistent. It then follows that this value $v_i$ must be unique among consistent data. By the arguments from case 1, we need it to be that $\max_{N \in \mathcal{N}_i^+ \cap \mathcal{N}_i^-}p_i^N \leq \min_{N \in \mathcal{N}_i^+ \cap \mathcal{N}_i^-}p_i^N$. Further, it must also be the case that $\max_{N \in \mathcal{N}_i^+ \cap \mathcal{N}_i^-}p_i^N \leq p_i^= \leq \min_{N \in \mathcal{N}_i^+ \cap \mathcal{N}_i^-}p_i^N$. However, $N^* \in \mathcal{N}_i^+\cap \mathcal{N}_i^-$, so $\max_{N \in \mathcal{N}_i^+ \cap \mathcal{N}_i^-}p_i^N \geq p_i^= \geq \min_{N \in \mathcal{N}_i^+ \cap \mathcal{N}_i^-}p_i^N$. It then follows that the three conditions hold if and only if we the dataset is consistent with GLM.
    \end{enumerate}
\end{proof}

\subsection{Proof of Proposition \ref{Prop:genidentification}}
\begin{proof}
    This follows immediately from Lemma \ref{inverseconelemma} and Theorem \ref{Thm:GenChar}.
\end{proof}

\subsection{Proof of Corollary \ref{cor:PointIdent}}
\begin{proof}
    By consistency with the linear-in-means model, $co^{-1}(\Delta(p^{N_j}),p_i^{N_j})$ and $co^{-1}(\Delta(p^{N_k}),p_i^{N_k})$ intersect. By $N_j$ and $N_k$ being binary, $co^{-1}(\Delta(p^{N_j}),p_i^{N_j})$ and $co^{-1}(\Delta(p^{N_k}),p_i^{N_k})$ are one-dimensional rays, and so they intersect once or infinitely often. By linear independence of $(p_i^{N_j}-p_j^{N_j})$ and $(p_i^{N_k}-p_k^{N_k})$ and by the definition of $co^{-1}(\Delta(p^{N_j}),p_i^{N_j})$ and $co^{-1}(\Delta(p^{N_k}),p_i^{N_k})$ , these two sets intersect at a single point. By Proposition \ref{Prop:genidentification}, this single point of intersection is exactly our identified set for $v_i$.
\end{proof}

\subsection{Proof of Theorem \ref{thm:pointIdent} and Theorem \ref{thm:NegPointIdent}}
In what follows, for two sets $A,B\subseteq \mathbb{R}^n$, let $A+B=\{x|x=a+b,a\in A, b\in B\}$. When $A$ is singleton, i.e. $A=\{x\}$, we utilize the abuse of notation $x+B$ to mean $\{x\}+B$. We begin by stating and proving a standard result in linear and affine algebra.
\begin{theorem}\label{thm:aff}
    Suppose that $U$ and $V$ are linear subspaces of $\mathbb{R}^n$ and let $u,v\in \mathbb{R}^n$ be vectors. If $(u+U)\bigcap (v+V) \neq \emptyset$, then there exists $w\in \mathbb{R}^n$ such that $(u+U)\bigcap (v+V) = w+(U \cap V)$.
\end{theorem}
\begin{proof}
    We begin with a preliminary lemma. \begin{lemma}
        Suppose that $V\subseteq \mathbb{R}^n$ is a linear subspace, $U\subseteq V$ is also a linear subspace, and $x,y \in V$. The following are equivalent.
        \begin{enumerate}
            \item $x-y \in U$
            \item $x+U=y+U$
            \item $(x+U)\cap(y+U)\neq \emptyset$
        \end{enumerate}
    \end{lemma}
    \begin{proof}
        We begin by proving the equivalence between the first and third conditions. Suppose that $x-y \in U$. This and the fact that $0 \in U$ implies that $x,y \in y +U$. Similarly, it is also the case that $y,x \in x +U$. We then have that $x,y \in (x+U) \cap (y+U)$, which proves our third condition. Now suppose that $(x+U)\cap(y+U)\neq \emptyset$ and let $z \in (x+U)\cap(y+U)$. This means that, for $u',u'' \in U$, $z=x+u'=y+u''$. It then follows that $x-y=u''-u'\in U$ which gives us our first condition.

        We now prove the equivalence between the second condition and the other two conditions. Let $z \in (x+U)\cap(y+U)$. This means that, for $u' \in U$, $z= x+u'$. Since $(y-x)\in U$ by the equivalence of the first and third conditions, we have that $z+(y-x)=x+(y-x)+u' = y+u'\in x+U$. This means that, for all $u'\in U$, $y+u' \in x+U$. By analogous argument, for all $u' \in U$, $x + u' \in y+U$. This gives us $x+U=y+U$, our second condition. Now suppose that $z\in (x+U) = (y+U)$. This means that, for some $u',u''\in U$, $z= x + u'= y+u''$. It then follows that $x-y=u''-u' \in U$, giving us our first condition. Thus all three conditions are equivalent.
    \end{proof}
    We now return to the proof of Theorem \ref{thm:aff}. Let $x \in (u+U)\cap (v+V)$. This means that, for $u' \in U$ and $v' \in V$, $x= u + u'=v+v'$. This gives us that $x-u=u'\in U$. By our lemma, we get that $x+U=u+U$. By analogous argument, we get that $v+V=x+V$. Now suppose that $y \in (x+U) \cap (x+V)$. This means that, for $u'' \in U$ and $v''\in V$, we have that $y=x+u''=x+v''$, but it must be that $u''=v''\in U \cap V$. This gives us $(u+U)\cap (v+V)=(x+U)\cap(x+V) \subseteq x+(U \cap V)$. Now suppose that $z \in x+(U\cap V)$. This means, for some $w \in U \cap V$, $z=x+w$, which further implies that $z=x+w\in x+U$ and $z=x+w\in x +V$. This gives us that $x+(U\cap V)\subseteq (x+U)\cap(x+V)=(u+U)\cap(v+V)$, which concludes our proof.
\end{proof}

We now turn to the proof of Theorem \ref{thm:pointIdent} and Theorem \ref{thm:NegPointIdent}.
\begin{proof}
    Fix agent $i$ and suppose that the data is consistent with GLM. Further, suppose that for each non-zero vector $p$ in the span of $\{p_i^N-p_j^N\}_{N \in \mathcal{N}_i,j \in N\setminus i}$, there is some group $M \in \mathcal{N}_i$ such that $p$ is not in the span of $\{p_i^M-p_j^M\}_{j \in M \setminus  i}$. For each group $N\in \mathcal{N}_i$, consider the linear subspace $U^N=span(\{p_i^{N}-p_j^{N}\}_{j \in N\setminus i})$. By construction, $(\sum_{N\in \mathcal{N}_i}U^N)= span(\{p_i^N-p_j^N\}_{N \in \mathcal{N}_i,j \in N\setminus i})=V.$ Fix a $U^{N'}$. By assumption (i.e. the condition in the theorem), for each non-zero vector $u \in U^{N'}$, there is some $M\in \mathcal{N}_i$ such that $u \not \in U^M$. It follows that $\bigcap_{N \in \mathcal{N}_i}U^N=\{0\}$. Now observe that each $co^{-1}(\Delta(p^{N}),p_i^{N})$ is a subset of the affine subspace $p_i^N+U^N$. Now consider $\bigcap_{N \in \mathcal{N}_i} (p_i^N+U^N)$. By the assumption that the data is consistent with GLM, $\bigcap_{N \in \mathcal{N}_i} (p_i^N+U^N)$ is non-empty as $\bigcap_{N \in \mathcal{N}_i}co^{-1}(\Delta(p^{N}),p_i^{N})$ is non-empty. By Theorem \ref{thm:aff}, this means that there exists some vector $v$ such that $v + (\bigcap_{N \in \mathcal{N}_i}U^N)= (\bigcap_{N \in \mathcal{N}_i}p_i^N + U^N)$. By earlier arguments, we know that $(\bigcap_{N \in \mathcal{N}_i}U^N)=\{0\}$, so we have that $(\bigcap_{N \in \mathcal{N}_i}p_i^N + U^N)= v + (\bigcap_{N \in \mathcal{N}_i}U^N)=\{v\}$. Since the data is consistent with GLM, $v \in \bigcap_{N \in \mathcal{N}_i}co^{-1}(\Delta(p^{N}),p_i^{N})$ and thus we have $\{v\}=\bigcap_{N \in \mathcal{N}_i}co^{-1}(\Delta(p^{N}),p_i^{N})$, giving us point identification of $v_i$.
\end{proof}
Observe that the proof of Theorem \ref{thm:NegPointIdent} proceeds exactly as the proof of Theorem \ref{thm:pointIdent} once we replace vectors of the form $p_i^N -p_j^N$ with vectors of the form $p_j^N$.
\subsection{Proof of Proposition \ref{Prop:PiPointIdent}}
\begin{proof}
    As $v_i$ and $\{p_k^N\}_{k \in N\setminus i}$ are affinely independent, any point in their convex hull can be written as a convex combination of these points with unique weights on each point. In the linear-in-means model, we have $p_i^N = \pi^N_i(i)v_i + \sum_{j\neq i}\pi^N_i(j)p^N_j$, and so $p_i^N$ is in the convex hull of these points and thus each $\pi^N_i(j)$ is uniquely pinned down.
\end{proof}

\subsection{Proof of Theorem \ref{Thm:LLMCharacterization}}

\begin{proof}

To begin observe that consistently weakly differentiates is satisfied if and only if there does not exist a collection of vectors $\{\alpha^N\}_{N \in \mathcal{N}_i}$ with $\alpha^N \in \mathbb{R}^X$ and $\sum_{N \in \mathcal{N}_i} \alpha^N\ll 0$ such that for all $N\in \mathcal{N}_i$, $\alpha^N \cdot p_i^N>0$ and for each $j\neq i$, $\sum_{N\in \mathcal{N}_i:j\in N}\alpha^N \cdot (p_i^N-p_j^N)\geq 0$. Now suppose that LLM is satisfied, and by means of contradiction, that there exist $\alpha^N$ as in the previous sentence. Drop dependence of $w_i$ on $i$ to simplify notation. We use the notation $w(N) = \sum_{j\in N}w(j)$.

Observe that \begin{align*}\sum_{N\in \mathcal{N}_i}w(N)\alpha^N\cdot p_i^N=&\sum_{N\in \mathcal{N}_i}\left(w(i)\alpha^N \cdot v_i + \sum_{j\in N\setminus i}w(j)\alpha^N \cdot p_j^N\right)\\
=&w(i)\left(\sum_{N\in \mathcal{N}_i}\alpha^N\right) \cdot v_i + \sum_{j\in\mathcal{A}\setminus i}w(j)\sum_{N\in \mathcal{N}_i:j\in N}\alpha^N \cdot p_j^N\\<&\sum_{j\in\mathcal{A}\setminus i}w(j)\sum_{N\in \mathcal{N}_i:j\in N}\alpha^N \cdot p_j^N\\ \leq& \sum_{j\in\mathcal{A}\setminus i}w(j)\sum_{N\in \mathcal{N}_i:j\in N}\alpha^N \cdot p_i^N\\=&\sum_{N\in \mathcal{N}_i}\sum_{j\in N\setminus i}w(j)\alpha^N \cdot p_i^N.\end{align*} 

Here, the strict inequality follows as $v_i \in \Delta(X)$, $w(i)>0$, and $\sum_{N\in \mathcal{N}_i}\alpha^N \ll 0$. The weak inequality follows as $\sum_{N\in \mathcal{N}_i:j\in N}\alpha^N (p_i^N-p_j^N)\geq 0$ and $w(j)\geq 0$.  The equalities are by definition and algebraic manipulation. Now, subtracting the right hand side of the string of inequalities from the left hand side, we obtain:  $$\sum_{N\in \mathcal{N}_i}w(i)\alpha^N \cdot p_i^N < 0.$$ This contradicts the facts that $w(i)>0$ and for each $N\in \mathcal{N}_i$, $\alpha^N \cdot p_i^N > 0$.

For the other direction, we want to find, for each $i$, numbers $f_i(x)\in\mathbb{R}$, and for each $j\neq i$, $w_i(j)\in\mathbb{R}$ and finally for each $N$ for which $i\in N$, $\lambda_i(N)$ such that the following equations are satisfied:

\begin{enumerate}
\item $f_i(x) \geq 0$ for all $x$
\item $0<\sum_x f_i(x)$
\item $w_i(j) \geq 0$ for all $j\neq i$
\item\label{eq:bigequation} $f_i(x) + \lambda_i(N)p_i^N(x) + \sum_{j\in N\setminus i}w_i(j)p_j^N(x) = 0$ for all $x$ and all $N$ for which $i\in N$
\end{enumerate}

If we find such numbers, define the Luce weights for agent $i$ as $w(i) = \sum_x f_i(x)>0$, $w(j) = w_i(j)\geq 0$ and define $v_i(x) = \frac{f_i(x)}{\sum_x f_i(x)}$.  Observe then that by equation~\ref{eq:bigequation}:

$w(i)v_i(x) + \lambda_i(N)p_i^N(x) +\sum_{j\in N\setminus i}w(j)p_j^N(x)=0$, so that $$-\lambda_i(N)p_i^N(x) = w(i)v_i(x)+\sum_{j\in N \setminus i}w(j)p_j^N(x).$$

Since we know $\sum_x v_i(x)=1$ (by definition), $\sum_x p_i^N(x) = 1$ and $\sum_x p_j^N(x) =1$ (by assumption), it follows automatically that $-\lambda_i(N)=w(N)$ and we have shown that this is a linear formulation of Luce.

The dual of this system, a Theorem of the Alternative, for example Motzkin's Theorem, \citep{mangasarian} p. 28  implies the existence of $\beta^N\in \mathbb{R}^X$ for each $N\in \mathcal{N}_i$ for which
\begin{enumerate}
\item For all $N\in\mathcal{N}$ for which $i\in N$, $\beta^N \cdot p_i^N = 0$.
\item For all $j\in\mathcal{A}\setminus i$, $\sum_{N\in \mathcal{N}_i:j\in N}\beta^N\cdot p_j^N\leq 0$.
\item $\sum_{N\in \mathcal{N}_i}\beta^N\ll 0$.
\end{enumerate}

Now, owing to finiteness of $X$, we know that there exists some $b<0$ such that for each $x\in X$, $\sum_{N\in \mathcal{N}_i}\beta^N(x) < b$.  Define $\alpha^N = \beta^N -\frac{b}{|\mathcal{N}_i|}\mathbf{1}$.  Observe that $\alpha^N \cdot p_i^N = \beta^N \cdot p_i^N - \frac{b}{|\mathcal{N}_i|}=\frac{-b}{|\mathcal{N}_i|}>0$.  Observe that $\sum_{N\in \mathcal{N}_i:j\in N}\alpha^N \cdot p_j^N \leq \frac{-b|\{N\in \mathcal{N}_i:j\in N\}|}{|\mathcal{N}_i|}=\sum_{N\in \mathcal{N}_i:j\in N}\alpha^N \cdot p_i^N$, so that $\sum_{N\in \mathcal{N}_i:j\in N}\alpha^N \cdot (p_i^N-p_j^N) \geq 0$.  Finally, $\sum_{N\in \mathcal{N}_i}\alpha^N = (\sum_{N\in \mathcal{N}_i}\beta^N )-b\mathbf{1}\ll 0$.

\end{proof}

\section{GLM Results when $\pi_i^N(i) \geq 0$}\label{app:GLMExtension}
In this section we consider an extension of GLM where we now allow $\pi_i^N(i) \geq 0$ (allowing for equality). We call this version of the model GLM$^*$. Our goal is to discuss and prove how our main results on GLM extend to GLM$^*$. Recall that for an agent $i$, $\mathcal{N}_i^{ext}$ corresponds to the set of groups $N$ containing $i$ with $p_i^N \not \in \Delta(p_{-i}^N)$. Observe that in GLM$^*$, whenever $p_i^N$ lies in $\Delta(p_{-i}^N)$, $p_i^N$ is consistent with GLM$^*$ as it can be written as a convex combination of $p_{-i}^N$. This motivates the following.

\begin{definition}
     We say that a set of vectors $\{b^N\}_{N \in \mathcal{N}_i}$ with $b^N \in \mathbb{R}^X$ for each $N \in \mathcal{N}_i$, \textbf{ext behaviorally differentiates} agent i if the following are satisfied.
     \begin{enumerate}
         \item (ext-Alignment) For each $N \in \mathcal{N}_i^{ext}$, $b^N \cdot p_i^N > 0$.
         \item (ext-Differentiation) For each $N \in \mathcal{N}_i^{ext}$ and each $j \in N \setminus i$, $b^N\cdot (p_i^N - p_j^N) \geq 0$.
     \end{enumerate}
\end{definition}
\begin{definition}
    We say that a dataset $\{p^N\}_{N \in \mathcal{N}}$ \textbf{ext consistently differentiates} agent $i$ if every $\{b^N\}_{N\in \mathcal{N}_i}$ which behaviorally differentiates agent $i$ does not satisfy $\sum_{N \in \mathcal{N}_i^{ext}}b^N \ll 0$.
\end{definition}

\begin{theorem}\label{Thm:GenCharExtended}
For a dataset $\{p^N\}_{N \in \mathcal{N}}$, the following are equivalent.
\begin{enumerate}
    \item $\{p^N\}_{N \in \mathcal{N}}$ is consistent with GLM$^*$.
    \item For every $i \in \mathcal{A}$, the collection of sets $\{co^{-1}(\Delta(p^N),p_i^N)\}_{N \in \mathcal{N}_i^{ext}}$ has a point of mutual intersection.
    \item For all agents $i \in \mathcal{A}$, $\{p^N\}_{N \in \mathcal{N}}$ ext consistently differentiates agent $i$.
\end{enumerate}
\end{theorem}

Before proving Theorem \ref{Thm:GenCharExtended}, we first prove a preliminary lemma.

\begin{lemma}\label{inverseconelemmaExtended}
    Suppose $p_i^N \not \in \Delta(p_{-i}^N)$. Then $p_i^N = \gamma_i v_i + \sum_{j \in N \setminus i} \gamma_j p_J^N$ for $v_i \in \Delta(X)$ with $\gamma_j \geq 0$ and $\sum_{j \in N}\gamma_j=1$ if and only if $v_i \in co^{-1}(\Delta(p^N),p_i^N)$.
\end{lemma}

\begin{proof}
    Suppose that $p_i = \gamma_i v_i + \sum_{j \in N \setminus i} \gamma_j p_J^N$ for $v_i \in \Delta(X)$ with $\gamma_j \geq 0$ and $\sum_{j \in N}\gamma_j=1$. Since $p_i^N \not \in \Delta(p_{-i}^N)$, it must be the case that $\gamma_i>0$ and so $0 < \gamma_i \leq 1$. By doing basic algebra we can recover that $v_i = \frac{1}{\gamma_i} p_i^N + \sum_{j \in N \setminus i}\frac{-\gamma_j}{\gamma_i}p_j^N$. Since $\gamma_j \geq 0$ and $\sum_{j \in N}\gamma_j=1$, $\frac{-\gamma_j}{\gamma_i} \leq 0$ and $\frac{1}{\gamma_i} + \sum_{j \in N \setminus i}\frac{-\gamma_j}{\gamma_i}=1$ and so $v_i \in co^{-1}(\Delta(p^N),p_i^N)$.

    Now suppose that $v_i \in co^{-1}(\Delta(p^N),p_i^N)$. This tells us that $v_i = \sum_{j \in N}\gamma_j p_j^N$ with $\gamma_j \leq 0$ $\forall j \in N\setminus i$ and $\sum_{j \in N}\gamma_j = 1$. It follows that $\gamma_i \geq 1$. By basic algebra, we can solve for $p_i$ and get $p_i=\frac{1}{\gamma_i} v_i + \sum_{j \in N \setminus i}\frac{-\gamma_j}{\gamma_i}p_j^N$. Since $\gamma_j \geq 0$ for $j \in N\setminus i$ and $\sum_{j \in N}\gamma_j=1$, $\frac{-\gamma_j}{\gamma_i} \geq 0$ and $\frac{1}{\gamma_i} + \sum_{j \in N \setminus i}\frac{-\gamma_j}{\gamma_i}=1$ and so $p_i = \pi_i v_i + \sum_{j \in N \setminus i} \pi_j p_J^N$ for $v_i \in \Delta(X)$ with $\pi_j \geq 0$ and $\sum_{j \in N}\pi_j=1$.
\end{proof}

We now proceed to prove Theorem \ref{Thm:GenCharExtended}.

\begin{proof}
    The equivalence between (1) and (2) follows from Lemma \ref{inverseconelemmaExtended} and our discussion at the start of this section. To see the equivalence between (1) and (3), observe that, in the proof of Theorem \ref{Thm:GenChar}, the arguments from the proof of (1)$\implies$(3) go through replacing $\mathcal{N}_i$ with $\mathcal{N}_I^{ext}$ and observing that once $\pi_i^N \not \in \Delta(p_{-i}^N)$ any coefficient preceding $\pi_i^N$ will be non-zero (see the first two sentences of the proof of Lemma \ref{inverseconelemmaExtended}). To show (3) implies (1), observe that the arguments from (3)$\implies$(1) from the proof of Theorem \ref{Thm:GenChar} hold when we try and prove the separation of $\Delta(X)$ and $\{co^{-1}(\Delta(p^N),p_i^N)\}_{N \in \mathcal{N}_i^{ext}}$.
\end{proof}

For the one-dimensional case, let $\mathcal{N}_i^{-*} \subseteq \mathcal{N}_i$ denote the set of groups $N$ satisfying $p_i^N < p_j^N$ for each $j \in N \setminus i$. Similarly, let $\mathcal{N}_i^{+*} \subseteq \mathcal{N}_i$ denote the set of groups $N$ satisfying $p_i^N > p_j^N$ for each $j \in N \setminus i$.

\begin{corollary}\label{cor:1DimCharExt}
    In the one-dimensional case, a dataset $\{p^N\}_{N \in \mathcal{N}}$ is consistent with GLM$^*$ if and only if, for all $i\in\mathcal{A}$, $\min_{N \in \mathcal{N}_i^{-*}} p_i^N \geq \max_{\mathcal{N}_i^{+*}} p_i^N$ when both $\mathcal{N}_i^{-*}$ and $\mathcal{N}_i^{+*}$ are non-empty.
\end{corollary}

\begin{proof}
    Observe that $\mathcal{N}_i^{-*} \cup \mathcal{N}_i^{+*} = \mathcal{N}_i^{ext}$ in the one dimension case. Further, by the arguments in the proof of Corollary \ref{cor:1DimChar}, $co^{-1}(\Delta(p^N),p_i^N) = [0,p_i^N]$ when $N \in \mathcal{N}_i^{-*}$ and $co^{-1}(\Delta(p^N),p_i^N) = [p_i^N,1]$ when $N \in \mathcal{N}_i^{+*}$. It then follows from Theorem \ref{Thm:GenCharExtended} that our corollary holds.
\end{proof}

\begin{proposition}\label{Prop:genidentification*}
    In GLM$^*$, the sharp identified set for $v_i$ is given by $\bigcap_{N \in \mathcal{N}_i^{ext}} co^{-1}(\Delta(p^N),p_i^N)$.
\end{proposition}

\begin{proof}
    This is an immediate consequence of Lemma \ref{inverseconelemmaExtended} and Theorem \ref{Thm:GenCharExtended}.
\end{proof}

All of the other identification results for GLM carry through to GLM$^*$ as written.

\section{Relation Between Theorem \ref{Thm:GenChar} and \citet{samet1998common}}\label{app:Samet}

In this appendix, we discuss the relation between \citet{samet1998common} and our characterizations of GLM and GLM$^*$. The main theorem of \citet{samet1998common} provides a linear program that characterizes when a collection of compact convex sets fail to have a point of mutual intersection. In light of Theorems \ref{Thm:GenChar} and \ref{Thm:GenCharExtended}, the result of \citet{samet1998common} can be used as an alternative characterization for GLM and GLM$^*$.

\begin{theorem}\label{thm:GLMSamet}
The following are equivalent.
\begin{enumerate}
    \item $\{p^N\}_{N \in \mathcal{N}}$ is consistent with the general linear-in-means model.
    \item For every $i \in \mathcal{A}$ there does not exist a collection of vectors $\{\alpha^N\}_{N \in \mathcal{N}_i}$ with $\alpha^N \in \mathbb{R}^X$ and $\sum_{N \in \mathcal{N}_i} \alpha^N=0$ such that, for all $N \in \mathcal{N}_i$, $\alpha^N \cdot v >0$ for each $v \in co^{-1}(\Delta(p^N),p_i^N) $.
\end{enumerate}
\end{theorem}

Theorem \ref{thm:GLMSamet} immediately follows from condition (2) of Theorem \ref{Thm:GenChar}, the main result of \citet{samet1998common}, and the observation that each $co^{-1}(\Delta(p^N),p_i^N)$ is compact and convex. Part of the original motivation for \citet{samet1998common} was the no interim trade result of \citet{morris1994trade}. In the setting of \citet{samet1998common} and \citet{morris1994trade}, $\{\alpha^N\}_{N \in \mathcal{N}_i}$ corresponds to a trading scheme, the equality condition corresponds to a feasibility condition, and the inequality condition corresponds to an individual rationality condition. 

We now interpret Theorem \ref{thm:GLMSamet} in the context of a no trade story. Suppose we have a collection of outside observers. We have one observer for each group $N$ in $\mathcal{N}_i$ and we index each observer by their corresponding group $N$. These observers are risk neutral but ambiguity averse (in the max-min sense of \citet{gilboa1989maxmin}) and consider state contingent trades. In this setting, a state corresponds to an alternative $x$ and the frequency of state $x$ corresponds to the frequency with which agent $i$ would choose $x$ in isolation, $v_i(x)$. In this sense, $\{\alpha^N\}_{N \in \mathcal{N}_i}$ corresponds to a trade scheme in the language of \citet{morris1994trade} where $\alpha^N(x)$ corresponds to the payout to observer $N$ when agent $i$ chooses $x$ in isolation. Each observer knows that agent $i$'s choices follow from Equation \ref{EQ:linInMeans}. The information available to observer $N$ is $p^N$, the set of choices made in group $N$. From this, observer $N$ is able to form a set of beliefs about the choices of agent $i$. This set of beliefs corresponds to $co^{-1}(\Delta(p^N),p_i^N)$. The condition $\sum_{N \in \mathcal{N}_i} \alpha^N=0$ corresponds to a trading scheme being balanced and $\alpha^N \cdot v >0$ for each $v \in co^{-1}(\Delta(p^N),p_i^N)$ corresponds to the trading scheme being profitable in expectation for every belief in observer $N$'s set of beliefs. With this interpretation, Theorem \ref{thm:GLMSamet} says that there is no trade between these outside observers if and only if the dataset is consistent with GLM. In other words, there is no trade if and only if each agent's ideal point $v_i$ is common across their groups.

Before moving on, we also note that condition (2) from Theorem \ref{thm:GLMSamet} can be simplified somewhat. Since $co^{-1}(\Delta(p^N),p_i^N)$ is a convex set and $\alpha^N \cdot v > 0$ is a linear constraint, instead of checking $\alpha^N \cdot v > 0 $ for each $v \in co^{-1}(\Delta(p^N),p_i^N)$, we can simply check $\alpha^N \cdot v > 0$ for the extreme points of $co^{-1}(\Delta(p^N),p_i^N)$. Since $co^{-1}(\Delta(p^N),p_i^N)$ is polyhedral, this is a finite process. Finally, we state the analogue of Theorem \ref{thm:GLMSamet} for GLM$^*$.

\begin{theorem}\label{thm:GLMStarSamet}
The following are equivalent.
\begin{enumerate}
    \item $\{p^N\}_{N \in \mathcal{N}}$ is consistent with GLM$^*$.
    \item For every $i \in \mathcal{A}$ there does not exist a collection of vectors $\{\alpha^N\}_{N \in \mathcal{N}_i^{ext}}$ with $\alpha^N \in \mathbb{R}^X$ and $\sum_{N \in \mathcal{N}_i^{ext}} \alpha^N=0$ such that, for all $N \in \mathcal{N}_i^{ext}$, $\alpha^N \cdot v >0$ for each $v \in co^{-1}(\Delta(p^N),p_i^N) $.
\end{enumerate}
\end{theorem}

Theorem \ref{thm:GLMStarSamet} immediately follows from condition (2) of Theorem \ref{Thm:GenCharExtended}, the main result of \citet{samet1998common}, and the observation that each $co^{-1}(\Delta(p^N),p_i^N)$ is compact and convex.

\section{Proofs of Results from Section \ref{Sec:uniLuce}}\label{App:ULM}
All of the results from Section \ref{Sec:uniLuce} follow from an application of Lemma \ref{inverseconelemma} to Equation \ref{EQ:uniLinInMeans}, which recovers Equation \ref{EQ:UniInverseCone}, and an application of the corresponding results for GLM applied to $co^{-1}(\bar{p}_{-i}^N,p_i^N)$, the set of feasible ideal points for agent $i$, as is developed in Equation \ref{EQ:UniInverseCone}.

\end{document}